\documentclass[12pt]{article}

\pdfoutput=1

\usepackage[top=80pt,bottom=85pt,left=85pt,right=85pt]{geometry}
\usepackage{graphicx,color,subfigure}
\usepackage{float}
\usepackage{cite}
\usepackage[debug,pageanchor=false]{hyperref}
\definecolor{link}{rgb}{.8,.15,.1}
\hypersetup{colorlinks=true,linkcolor=link,citecolor=link,urlcolor=link,linktocpage}
\usepackage{tikz}
\usetikzlibrary{decorations.pathreplacing}
\usetikzlibrary{positioning}
\usepackage{setspace,verbatim} 
\usepackage{amsmath,amsfonts,amssymb,amsthm}

\newcommand{\R}{\mathbb{R}}

\newcommand{\zz}{\mathbb{Z}}

\newcommand{\del}{\partial}

\newcommand{\p}{p}
\newcommand{\z}{z}
\renewcommand{\k}{k}
\newcommand{\nn}{\nonumber}

\makeatletter
\@addtoreset{equation}{section}
\makeatother

\begin{document}

\begin{titlepage}

\begin{flushright} \small
UUITP-12/17
\end{flushright}

\begin{center}

\noindent

{\Large \bf{AdS$_5$ compactifications with punctures\\ in massive IIA supergravity}}

\bigskip\medskip

Ibrahima Bah$^{1,2}$, Achilleas Passias$^3$ and Alessandro Tomasiello$^4$\\

\bigskip\medskip
{\small

$^1$Department of Physics, University of California, San Diego, La Jolla, CA 92093 USA \\
\vspace{.3cm}
$^2$ Department of Physics and Astronomy, Johns Hopkins University,\\ 3400 North Charles Street, Baltimore, MD 21218, USA
\vspace{.3cm}
\\
$^3$Department of Physics and Astronomy, Uppsala University,\\
Box 516, SE-75120 Uppsala, Sweden
\\	
\vspace{.3cm}
$^4$Dipartimento di Fisica, Universit\`a di Milano--Bicocca, \\ Piazza della Scienza 3, I-20126 Milano, Italy \\ and \\ INFN, sezione di Milano--Bicocca
	
}

\vskip .5cm 
{\small \tt iboubah@jhu.edu, achilleas.passias@physics.uu.se , alessandro.tomasiello@unimib.it}
\vskip .9cm 
     	{\bf Abstract }

\vskip .1in
\end{center}

\noindent
We find AdS$_5$ solutions holographically dual to compactifications of six-dimensional ${\cal N}=(1,0)$ supersymmetric field theories on Riemann surfaces with punctures. We simplify a previous analysis of supersymmetric AdS$_5$ IIA solutions, and with a suitable Ansatz we find explicit solutions organized in three classes, where an O8--D8 stack, D6- and D4-branes are simultaneously present, localized and partially localized. The D4-branes are smeared over the Riemann surface and this is interpreted as the presence of a uniform distribution of punctures.  For the first class we identify the corresponding six-dimensional theory as an E-string theory coupled to a quiver gauge theory.  The second class of solutions lacks D6-branes and its central charge scales as $n^{5/2}$,  suggesting a five-dimensional origin for the dual field theory.  The last class has elements of the previous two.

\noindent

\vfill
\eject

\end{titlepage}

\tableofcontents

\section{Introduction} 
\label{sec:intro}

An interesting way of producing a conformal field theory is by reducing a higher-dimensional one. Among four-dimensional superconformal field theories, there is a vast number of examples that come from compactifying six-dimensional theories. This has been demonstrated for the ${\cal N}=(2,0)$ theory living on a stack of M5-branes: upon compactification on Riemann surfaces, it produces interesting ${\cal N}=2$ theories in four dimensions, known as ``class $\mathcal{S}$'' theories. This can be understood field-theoretically \cite{witten-D4,gaiotto,Gaiotto:2009hg} and from a holographic perspective \cite{maldacena-nunez,gaiotto-maldacena}. 
It is also possible to compactify the ${\cal N}=(2,0)$ theory in a more complicated way, so as to produce ${\cal N}=1$ theories \cite{maldacena-nunez,bah-beem-bobev-wecht,bah-bobev,agarwal-bah-maruyoshi-song,bah-D6}.

In six dimensions,  string theory also suggests the existence of a much larger class of superconformal field theories with ${\cal N}=(1,0)$ supersymmetry; see for example \cite{seiberg-witten-6d,seiberg-6d,intriligator-6d,intriligator-6d-II,hanany-zaffaroni-6d,brunner-karch,blum-intriligator}. Recently, certain classifications have been proposed, based on anomalies and supersymmetry \cite{bhardwaj} and on F-theory \cite{heckman-morrison-rudelius-vafa}. Moreover, a holographic description has been found \cite{afrt,gaiotto-t-6d,10letter,cremonesi-t} for a large class of such theories \cite{hanany-zaffaroni-6d,brunner-karch}. These theories have an effective description in terms of a chain of unitary gauge groups, coupled to tensor multiplets and hypermultiplets. They are engineered by an NS5--D6--D8 brane system. Their holographic duals are AdS$_7\times M_3$ solutions with D8/D6-brane sources, where $M_3$ has the topology of the three-sphere $S^3$. 

It is natural then to ask whether these six-dimensional ${\cal N}=(1,0)$ theories also produce superconformal field theories (SCFTs) upon compactification on Riemann surfaces $\Sigma$. In \cite{afpt,prt} it was shown holographically that by compactifying the class in \cite{afrt} on $\Sigma$ of genus $g>1$ one indeed obtains SCFTs with ${\cal N}=1$ supersymmetry in four dimensions.\footnote{The particular case $g=1$ was also considered for more general theories in   \cite{delzotto-vafa-xie,ohmori-shimizu-tachikawa-yonekura-T2,ohmori-shimizu-tachikawa-yonekura-S1T2,Bah:2017gph}.} A next step would be to understand these theories more directly. Although they are not expected to admit a Lagrangian description, they might be amenable to a decomposition similar to the one obtained for the class $\mathcal{S}$ theories in \cite{gaiotto,Gaiotto:2009hg}. There it was shown how to associate a ``generalized quiver'' description to a pair-of-pants decomposition of a Riemann surface $\Sigma$, namely to a choice of representation of $\Sigma$ as a union of three-punctured spheres and tubes. To each three-punctured sphere one associates a class of theories that correspond to compactifying the ${\cal N}=(2,0)$  theory in presence of codimension-2 defects of various types. The theories associated to $\Sigma$ can be constructed by gluing three-punctured spheres \cite{gaiotto,Gaiotto:2009hg}. In \cite{gaiotto-maldacena} the holographic dual for this construction was described.  

Obtaining a similar picture for ${\cal N}=(1,0)$ theories compactified on $\Sigma$ would greatly enlarge our knowledge of ${\cal N}=1$ SCFTs in four dimensions. So far, this has been attempted \cite{gaiotto-razamat,franco-hayashi-uranga,hanany-maruyoshi,razamat-vafa-zafrir,Bah:2017gph} for one particular ${\cal N}=(1,0)$ theory, namely the one describing a stack of M5-branes on top of a $\zz_k$ singularity. (This theory is also part of the class obtained in IIA in \cite{afrt}: it is the only one with vanishing Romans mass.) 

In this paper, we find AdS$_5$ solutions which are dual to compactifications on punctured Riemann surfaces of any genus, of certain ${\cal N}=(1,0)$ theories belonging to the family of \cite{afrt}. These ${\cal N}=(1,0)$ theories are obtained by coupling a six-dimensional quiver tail to the E-string theory of rank one (see figure \ref{fig:quiver}), a construction following \cite{hanany-zaffaroni-6d,brunner-karch}.  The E-string theory has a tensor multiplet but no gauge group, and an $E_8$ flavor symmetry. 
It first appeared as a description of an M5-brane near an M-theory $E_8$ wall \cite{Witten:1995gx,ganor-hanany,seiberg-witten-6d}, but since then, has often played a role in other contexts, for example in the various theories engineered in F-theory \cite{heckman-morrison-rudelius-vafa,heckman-morrison-vafa}. In our case, it appears because the six-dimensional theories we are compactifying have an AdS$_7$ dual which contains, in addition to more ``customary'' D6-branes, an O8--D8 stack.\footnote{This possibility was already implicitly present in \cite{10letter,afpt}, as a particular limit of the parameters, but was not noted at that time.} The corresponding brane diagram (see figure \ref{fig:O8-NS5-D6}) describes $N$ NS5-branes near the O8--D8 stack, in the presence of Romans mass; because of the latter, D6-branes are also created. The resulting theory can be thought of as an analogue with Romans mass, of the rank-$N$ E-string theory describing $N$ M5-branes near an $E_8$ wall. For this reason, we will refer to our six-dimensional models as ``massive E-string theories''. The holographic identification between the AdS$_7$ solution and these theories is also bolstered by an anomaly computation similar to \cite{cremonesi-t}, which yields a perfect match.

The AdS$_5$ solutions we find have in addition to the aforementioned D6-branes and O8--D8 stack, D4-branes, which represent the punctures on the Riemann surface. The D4-branes are extended along the AdS$_5$ and smeared over the Riemann surface $\Sigma$. The latter means that there are many punctures distributed over $\Sigma$. In the rest of the internal space which is topologically an $S^3$ (or rather, half-$S^3$, because of the presence of the O8-plane), the D4-branes are completely localized on a point on top of the O8-plane. 
Our solutions thus contain sources of three different kinds (O8--D8, D6, D4), almost all completely localized; this is an exceptionally complex set of localized ingredients. Identifying the various sources requires comparing the field behavior near them to the one known in flat space. This is nontrivial especially for the D4-branes inside the O8-plane, and is performed here with a variant of the analysis in \cite{youm}. 

We have checked the field theory interpretation of our solutions by computing the central charge:
\begin{equation}\label{eq:a}
	a=\frac{27}{32}\left(\frac15 (g-1) N^3 M^2 + \frac13 n N^2 M \right)\, ,
\end{equation} 
where $N$ is the number of NS5-branes, $M$ is the number of D6-branes, and $n$ is the number of D4-branes. 
The ratio $27/32$ is typical of ${\cal N}=1$ theories obtained by mass deforming $\mathcal{N}=2$ theories \cite{tachikawa-wecht}. In the parenthesis, the first term comes from the compactification of the massive E-string theory on a genus $g$ surface $\Sigma$. The $N^3 M^2$ behavior is typical of ${\cal N}=(1,0)$ theories engineered from $N$ NS5-branes and $M$ D6-branes \cite{gaiotto-t-6d,cremonesi-t}. The fact that it is proportional to $(g-1)$ is also standard \cite{gaiotto-maldacena,bah-beem-bobev-wecht}. The second term in (\ref{eq:a}) is, quite sensibly, proportional to the number of punctures $n$; the contribution of each puncture scales as $N^2 M$, suggesting that these are the analogue of the ``simple'' punctures of class $\mathcal{S}$ theories. Notice that for the case $\Sigma=S^2$ ($g=0$) the first term is negative; however, the second term is always positive enough to keep $a>0$, due to a lower bound $n\ge NM$ on the number of punctures in this case. (No such bound exists for $g>0$.) This again is in agreement with the intuition from class $\mathcal{S}$ theories, where a sphere cannot have too low a number of simple punctures. 

In order to find the solutions, we started from the classification of supersymmetric AdS$_5$ solutions of massive type IIA supergravity\cite{afpt}, but in a simpler reformulation which reduces the number of partial differential equations (PDEs) that characterize the classification. The latter acquire a form which is reminiscent of the Toda--Monge--Amp\`ere system in \cite{bah-D6}, itself a generalization of the Toda equation in \cite{lin-lunin-maldacena}. This observation can be useful for broader aims than the ones in this paper. We then used a separation of variables Ansatz inspired again in part by \cite{bah-D6}, in order to solve the PDEs. 

In the process of looking for our punctured compactifications, we have also found some superficially similar solutions, which however appear to represent rather different physics. In the $\Sigma=S^2$ case, if one varies a parameter beyond a certain value, one finds a solution without any D6-branes, and with zero NS--NS flux integral, indicating also an absence of NS5-branes before the near-horizon limit. Moreover, some of the D4-branes have now moved off the O8-plane. In this case $a$ scales as $n^{5/2}$, which is the same scaling as of the action of the AdS$_6$ solution of \cite{brandhuber-oz}, arising as the near-horizon geometry of a O8--D8--D4 system. This might suggest a relation between the two solutions, and hence a five-dimensional origin of the four-dimensional SCFTs dual to the AdS$_5$ solutions. For $g>1$, we also find a solution where the NS5-branes and D6-branes are still present, but all the D4-branes have moved off the O8-plane. In this case $a$ exhibits a mix of the behavior in (\ref{eq:a}) and of $a\sim n^{5/2}$. All these alternative solutions are intriguing, but we are not giving them an interpretation here.

The rest of this paper is organized as follows. In section \ref{sec:general_system} we present a reformulation of the classification of supersymmetric AdS$_5$ IIA solutions of \cite{afpt}. In section \ref{sec:new_solutions} we obtain our new analytic solutions and reduce their study to three classes, which we analyze in detail in section \ref{sec:cases}. The solutions of section \ref{sub:6d} are the ones which we interpret as punctured compactifications of six-dimensional ${\cal N}=(1,0)$  SCFTs. In section \ref{sec:6d} we review these SCFTs and their holographically dual AdS$_7$ solutions. 


\section{Supersymmetric AdS$_5$ solutions: general system} 
\label{sec:general_system}

We begin by presenting a refinement of the classification of supersymmetric AdS$_5$ solutions of massive type IIA supergravity worked out in \cite{afpt}. In particular, by introducing a new set of functions characterizing the solutions, we are able to reduce the number of partial differential equations that control the classification, as well as simplify their form. This simpler system of equations was inspired by the one obtained for M-theory AdS$_5$ solutions in \cite{bah-D6}, and its reduction along a flavor isometry to ten dimensions. The connection between the formalism presented here and the one in \cite{afpt} is summarized in appendix \ref{app:comparison}.

The metric for a general supersymmetric AdS$_5$ solution is 
\begin{align}
ds^2_{10} &= e^{2W} \left[ ds^2_{{\rm AdS}_5} + e^{2A} \left(dx_1^2 + dx_2^2 \right) + \frac{1}{3} e^{-6\lambda} ds^2_3 \right]\, , \label{5-5metric} \\ \nonumber
ds^2_{3} &= -\frac{4}{\partial_s D_s} \eta_\psi^2 - \partial_s \widetilde{D}_s\, ds^2  - 2 \partial_u D_s\, du ds  - \partial_u D_u\, du^2\, , \label{eq:3metric}
\end{align} 
where
\begin{equation}
\eta_\psi = d\psi -\frac{1}{2} \star_2 d_2 D_s
\end{equation}
and $\widetilde{D}_s = D_s - \frac{3}{2}\ln s$\footnote{The labeling of the potential $D_s$ is related to the one in \cite{bah-D6} as $(D_s,\widetilde{D}_s) \to (\widetilde{D}_s,D_s)$.}.
The Hodge star operator $\star_2$\footnote{The convention for its action is $\star_2 dx_1 =dx_2$ and $\star_2 dx_2 =-dx_1$.  This convention is opposite of the one used in \cite{bah-D6}.} and the exterior derivative $d_2$ are taken over the $(x_1,x_2)$ plane.  The functions $(D_u, D_s)$, which determine the solution, depend on $(x_i, s,u)$. The rest of the functions appearing in the metric are given in terms of these as follows:
\begin{equation}
e^{-6\lambda} = \frac{1}{8s} \frac{\det(h)}{\det(g)}\,, \qquad e^{4W} = - \frac{\partial_s D_s}{3\det(h)}\,, \qquad e^{2A} = \frac{1}{24} \det(h) e^{D_s}\,,
\end{equation} 
and
\begin{equation}
\begin{split}
	\det(g) &=\partial_u D_u \partial_s \widetilde{D}_s - \left(\partial_u D_s \right)^2\,, \\
	\det(h) &=\partial_u D_u \partial_s D_s - \left(\partial_u D_s \right)^2\,.  
\end{split}	
\end{equation}
The geometry has a U$(1)$ isometry whose orbits are parameterized by $\psi$. It is in fact a symmetry of the full solution and corresponds to the R-symmetry of the dual superconformal field theory.

The dilaton can be expressed as
\begin{equation}
e^{2\phi} =  e^{6W} e^{-6\lambda}\,.
\end{equation}
The NS--NS field strength $H$ is given by
\begin{align}
H &= d V_2 - \frac{1}{36 \sqrt2 s^{3/2}} du \wedge ds \wedge \eta_\psi + \frac{1}{72\sqrt{2s}} e^{D_s} \left(\det(h) du + \frac{3}{2s} \partial_u D_s ds \right) \wedge dx_1 \wedge dx_2\,, \nonumber \\
V_2 &\equiv \frac{1}{24 \sqrt{2s} } \partial_u e^{D_s} dx_1 \wedge dx_2 - \frac{1}{12\sqrt2 s^{3/2} \det(g)} \left(\partial_u D_u du + \partial_u D_s ds \right) \wedge \eta_\psi\,.
\end{align}  
The R--R field strengths read
\begin{align}
F_0 &= 36 \sqrt{2s} \frac{\partial_u \left(\partial_s D_u - \partial_u D_s \right)}{\partial_s D_s}\, ; \label{eq:F0}\\
F_2 &= dV_1 + \star_2 d_2 \left(\partial_u D_s - \partial_s D_u \right) \wedge ds  
+\left[\Delta_2 D_u - \partial_u \left(s \det(g) e^{D_s} \right)\right] dx_1 \wedge dx_2
\nonumber \\
&- F_0 \frac{1}{18 \sqrt{2s}} du \wedge \eta_\psi - F_0 \frac{1}{12\sqrt2 s^{3/2} \det(g)} \left(\partial_u D_u du + \partial_u D_s ds \right) \wedge \eta_\psi\, , \label{eq:F2} \\
V_1 &\equiv - \star_2 d_2 D_u - 2 \frac{\partial_u D_s}{\partial_s D_s} \eta_\psi\, ; \nonumber \\
F_4 &= \frac{1}{24\sqrt2} \frac{(\det(h))^2}{\det(g)} \frac{e^{D_s}}{\sqrt{s}\partial_s D_s} \nonumber \\
&
\times \Bigl\{ \Bigr.
\left[ \partial_u D_u \partial_s V_0 - \partial_u D_s \left( \partial_u V_0 + \tfrac{4}{3} \right) \right] du 
+ \left[\partial_u D_s \partial_s V_0 - \partial_s \widetilde{D}_s \left(\partial_u V_0 + \tfrac{4}{3}\right) \right] ds 
\Bigl. \Bigr\}  \nn \\
&\wedge d\psi \wedge dx_1 \wedge dx_2 \nonumber \\
&+ \frac{1}{24\sqrt2}\frac{(\det(h))^2}{\det(g)}\frac{1}{s^{3/2} \partial_sD_s} \star d_2 V_0 \wedge du \wedge ds \wedge D\psi\, ,\\
V_0 &\equiv \frac{2 \partial_u D_s}{\det(h)}\, . \nonumber
\end{align} 
In the above $\Delta_2$ is the Laplace operator $\Delta_2 = \partial^2_{x_1} + \partial^2_{x_2}$.

The Bianchi identity of the Romans mass $F_0$ sets it to a constant.  The one of the NS--NS field strength, $dH = 0$,  yields an equation for $D_s$:
\begin{equation}\label{eq:dH}
\Delta_2 D_s = \partial_s \left(s \det(g) e^{D_s} \right) + \frac{1}{24 \sqrt{2s}} F_0 \partial_s e^{D_s}\, . 
\end{equation}  
Given the above, the Bianchi identity of $F_2$, $dF_2 - F_0 H = 0$, becomes
\begin{equation}\label{eq:dF2}
\Delta_2\left(\partial_u D_u\right) = \partial_u^2 \left(s \det(g) e^{D_s} \right)  + \frac{1}{36 \sqrt{2s}} F_0 s \partial_s \left(\det(h) e^{D_s} \right)\, . 
\end{equation}
A potential extra integrability condition, obtained by acting on (\ref{eq:dH}) with $\partial_u^2$ and on (\ref{eq:dF2}) with $\partial_s$, is automatically satisfied upon using (\ref{eq:F0}). The Bianchi identity of $F_4$ is also automatically satisfied.

Summarizing, the solution is determined by $(D_u, D_s)$ which are subject to equations (\ref{eq:F0}), (\ref{eq:dH}) and (\ref{eq:dF2}).

\section{New solutions} 
\label{sec:new_solutions}

We will now look for new AdS$_5$ solutions, by using the system obtained in the previous section and introducing a suitable Ansatz. It involves the presence of a Riemann surface $\Sigma$, coordinatized by $x_1$, $x_2$. The remaining coordinates describe a three-manifold $M_3$, topologically fibred over $\Sigma$.

\subsection{Ansatz}

In this paper, we study solutions whose metric on the $(x_1,x_2)$ plane has constant sectional curvature. This requires the warp factor $A$ to be separable:
\begin{equation}
e^{2A} = f(s,u) e^{2A_0(x_1,x_2)} \label{sepA}
\end{equation} 
for some function $f(s,u)$. In order to satisfy the separability condition \eqref{sepA} we make an appropriate Ansatz for the potentials $(D_u,D_s)$.  The most general one is
\begin{equation}
D_s = F_s(s,u) + 2 A_0(x_1,x_2)\ , \qquad D_u = F_u(s,u) + 2 v \widetilde{A}_0(x_1,x_2)\, , \label{Dansatz}
\end{equation} where $v$ is an arbitrary parameter.

\subsubsection*{Riemann Surface}
In the Ansatz for the $D$ potentials, $A_0(x_1,x_2)$ is a solution of the equation
\begin{equation}\label{eq:liouville}
\left(\partial_{x_1}^2 + \partial_{x_2}^2 \right) A_0 + \kappa e^{2A_0} =0\,.  
\end{equation} 
The parameter $\kappa$ is restricted to the values $\{-1,0,1\}$, without loss of generality.  The three choices correspond to the hyperbolic space $H_2$, the torus $T^2$ and the two-sphere $S^2$ respectively.  The hyperbolic space $H_2$ can be replaced by the quotient $H_2/\Gamma$ to obtain a constant curvature Riemann surface of genus $g$. $\Gamma$ is a Fuchsian subgroup of the PSL$(2,\R)$ isometry group of $H_2$.   A representative solution to (\ref{eq:liouville}) is
\begin{equation}
e^{A_0} = \frac{2}{1+ \kappa \left(x_1^2 + x_2^2\right)}\, .  
\end{equation}  
The function $\widetilde{A}_0$ is given as
\begin{equation}
\widetilde{A}_0 = \begin{cases} - \left(x_1^2 + x_2^2\right) \quad &\mbox{for} \quad \kappa=0\, , \\
\kappa A_0 \quad &\mbox{for} \quad \kappa \neq 0\, .  \end{cases}
\end{equation} 

It is convenient to introduce the connection one-form, $V$, which is defined as
\begin{equation}
V =  -\frac{1}{2} \alpha_\kappa \star_2 d_2 \widetilde{A}_0 = \alpha_\kappa \frac{x_1 dx_2 - x_2 dx_1}{1+ \kappa \left(x_1^2 + x_2^2\right)}\,, \qquad 
\alpha_\kappa = \begin{cases} 
\displaystyle \frac{1}{4 \pi} \quad &\mbox{for} \quad \kappa=0 \, ;\vspace{.4cm} \\ 
\displaystyle \frac{\kappa}{1-g} \quad &\mbox{for} \quad \kappa \neq 0\, . 
\end{cases}  \label{alphak}
\end{equation}  
The normalizations are such that 
\begin{equation}
\Delta_2\widetilde{A}_0 = - e^{2A_0}\,, \qquad dV = \frac{\alpha_\kappa}{2} e^{2A_0} dx_1 \wedge dx_2\,, \qquad \int dV = 2\pi\, .  
\end{equation} 

The local metric on the Riemann surface is 
\begin{equation}
ds^2 \left(\Sigma_g\right) = e^{2A_0(x_1,x_2)} \left(dx_1^2 + dx_2^2 \right)\, . \label{sigmet}
\end{equation}  
The one-form dual to the $\partial_\psi$ Killing vector can be written as
\begin{equation}
\eta_\psi = d\psi - \frac{1}{2} \star_2 d_2 D_s = d\psi - 2 \left(g-1\right) V\, . \label{psiform}
\end{equation}

\subsubsection*{Ansatz for $(F_u,F_s)$}
When we plug the Ansatz for the $D$ potentials into the system of equations of section \ref{sec:general_system}, we obtain Monge--Amp\`ere equations for $(F_u,F_s)$. By studying how the known AdS$_5$ solutions in type IIA and 11D supergravity solve these, we refine our Ansatz as
\begin{align}
F_s(s,u) &= 2\nu + \ln f_1(t_1) + \ln f_2 (t_2)\,, \\
F_u(s,u) &= \frac{1}{\tau_1(s)} \ln f_1(t_1) + \frac{1}{\tau_2(s)} \ln f_2(t_2)\,,
\end{align}  
where $\nu$ is a constant and
\begin{equation}
t_1 = u+ b_1(s)\,, \qquad t_2 = \ell_0 u + b_2(s)\,, \qquad 
\end{equation}
with the parameter $\ell_0 \in \{0,1\}$.
Furthermore,
\begin{equation}
\tau_i  \equiv \partial_s b_i(s)\,, \qquad  \partial_s \frac{1}{\tau_i^2} = \frac{F_0}{18 \sqrt{2s}}\,.
\end{equation}

The aim is to find all solutions for $f_i$ in terms of the variables $t_i$. The solutions of \cite{bah-beem-bobev-wecht} are obtained by taking $f_i$ linear and by fixing $\ell_0=1$ and $F_0=0$.  The solutions of \cite{afpt} are also obtained by taking $f_i$ linear and $\ell_0=1$ but with non-zero $F_0$.  The solutions of \cite{gauntlett-martelli-sparks-waldram-M} can be obtained by fixing $\ell_0=0$ and by taking $f_1$ linear.

The class of solutions we study in this paper are
\begin{equation}
f_2 = c_2 + t_2\,, \qquad f_1 = c_1 +2 \kappa t_1 + f_0 \left(c_0 + t_1\right)^{1/3}\, \label{fsol}
\end{equation} where the set of parameters $(c_2,c_1,c_0,f_0)$ are integration constants.  In this class of solutions we have
\begin{equation}
\tau_1 = -\tau_2 = \left(\frac{18}{F_0 \sqrt{2s}} \right)^{1/2}\,.
\end{equation}  The $\tau$'s can be integrated for the $b$'s.  In order to study the solutions, we make a further coordinate transformation from $(t_1,t_2)$ to $(z,k)$:
\begin{equation}
c_0 + t_1 =2F_0 z^3, \qquad c_2 + t_2 =2 F_0 z^3 (1 - k^3) \,.  
\end{equation}  
We now write and study the solutions.

\subsection{Solutions}

The metric that follows from the solution \eqref{fsol} in the $(z,k)$ coordinates is 
\begin{align}
ds^2_{10} &= e^{2W} \left[ ds^2_{{\rm AdS}_5} - \frac{\p'}{9\z^2} ds^2_5 \right]\,,  \label{fullsol} \\
ds^2_{5}  &= ds^2(\Sigma_g) + \frac{3\z d\z^2}{\p} 
+ \frac{9\z^3}{3\p - \z\p'} 
\left[ \frac{\k d\k^2}{1-\k^3} + \frac{4}{3}\frac{(1-\k^3)\p}{3\p - \z\p'(1-\k^3)}\eta_\psi^2 \right]\,, \nn
\end{align}  
where
\begin{equation}\label{p(z)}
\p = (\z-\z_0)\left[ \kappa(\z^2 + \z_0 \z + \z_0^2) - 3\ell\z^2_1 \right]
\end{equation}
and a prime denotes differentiation with respect to $\z$. Explicitly,
\begin{equation}
\p' = 3(\kappa \z^2 - \ell \z_1^2)\,.
\end{equation}
The parameters $\z_0$, $\z_1$ and $\ell$ are real, and $\ell \in \{-1, 1\}$\footnote{These parameters are related to the ones in \eqref{fsol} as 
\begin{equation}
f_0 = -6 \left(2F_0\right)^{2/3} \ell z_1^2\,, 
\qquad
c_1-2 \kappa c_0 = 4z_0F_0\left(3\ell z_1^2 - \kappa z_0^2 \right)\,.
\end{equation}
}. Without loss of generality $\z_1$ is chosen to be non-negative. The parameter $\kappa$ is the curvature of the Riemann surface $\Sigma_g$ of genus $g$, with local metric given in \eqref{sigmet}.  The one-form $\eta_\psi$ is also given above in \eqref{psiform}.  

The warp factor is given by the expression
\begin{equation}
e^{4W} = \frac{\z}{\k} \frac{3\p - \z\p'(1-\k^3)}{-\p'}\, . 
\end{equation}

The metric on AdS$_5$ is taken to be of unit radius. A radius $L$ can be reinstated by rescaling 
\begin{equation}\label{eq:resc}
\z \to L^2 \z\,, \qquad \z_0 \to L^2 \z_0\,, \qquad \z_1 \to L^2 \z_1\,.
\end{equation}

Positivity of the metric requires
\begin{equation}\label{eq:positivity}
\z \p \geq 0 \,, \qquad -\p' \geq 0 \,,\qquad  0 \leq \k \leq 1\, .
\end{equation}  
The metric (and indeed the complete solution) is invariant under the simultaneous reflection $\z \to -\z$ and $\z_0 \to -\z_0$, and so we will restrict our study to $\z \geq 0$. 

The dilaton reads
\begin{equation}
e^{4\phi} = \frac{1}{(F_0)^4} \frac{1}{\z^3\k^5}
\frac{[ 3\p - \z\p'(1-\k^3) ]^3}{-\p'(3\p - \z\p')^2}\,. 
\end{equation}

Finally, we present the expressions of the fluxes in terms of potentials $B$ for the NS--NS flux and $C_1$, $C_3$ for the R--R fluxes:
\begin{equation}
H = dB \, , \qquad F_2 = dC_1 + F_0 B\, , \qquad F_4 = dC_3 + B \wedge F_2 - \frac{1}{2} F_0 B \wedge B \, ,
\end{equation}
subject to the gauge transformations
\begin{equation}
\delta B = d\Lambda_1\,, \qquad 
\delta C_1 = d \Lambda_0 - F_0 \Lambda_1\,, \qquad 
\delta C_3 = d\Lambda_2 + \frac{1}{2} F_0 d\Lambda_1 \wedge \Lambda_1  
- d\Lambda_1 \wedge C_1\, .
\end{equation}
Their expressions are
\begin{align}
\label{eq:B} 
B &= - \frac{2}{3} \frac{\z^2\p'}{3\p-\z\p'} d\k \wedge \eta_\psi
- \frac{\k}{9}\frac{\p'-\z\p''}{\z}{\rm vol}_{\Sigma_g}\,, \\
\label{eq:C1}
C_1 &=  \frac{2}{3} F_0 \frac{\k\z^2\p'(1-\k^3)}{3\p-\z\p'(1-\k^3)} \eta_\psi\,, \\
\label{eq:C3}
C_3 &= \frac{2}{9} F_0 \k^2 
\left[ \frac{\p'-\z\p''}{3\p-\z\p'(1-\k^3)}\p + \frac{1}{6}\z\p'' \right] \eta_\psi \wedge {\rm vol}_{\Sigma_g}  \,.
\end{align}
This gauge choice is particularly convenient for the purpose of presentation. As we will see, the above potentials are actually singular at certain loci. They will however be sufficient for computing the charges of the various sources in our solutions. We postpone a more careful treatment to section \ref{sub:flux}. 

\subsection{Regularity and brane sources}
\label{sub:special_points}

In this section we study the geometry in special regions where the $S^1$ parameterized by $\psi$ shrinks or where the metric is singular. The latter regions are (i) $\k=0$, (ii) $\z=0$ and (iii)  $\p'=0$. We will demonstrate that these singularities correspond to brane sources.  This will be one of the main results in this paper.  

The regions where the $S^1$ shrinks are (i) $\p=0$ and (ii) $\k=1$.  We will begin by examining these.  

First we consider a region where $p=0$ and $p' \neq 0$.  Let $z_r$ be a single root of $p$.  We expand $p$  around $z_r$, $p(z) = p'(z_r) (z-z_r) + O\left((z-z_r)^2\right)$, and introduce a new coordinate, $\varrho$, as $\varrho^2 = \frac{4}{3} (1-z/z_r)$.
The metric in the region $z=z_r$ takes the form
\begin{equation}
ds^2_{10} \sim z_r \sqrt{\frac{1-k^3}{k}} \left[ds^2_{{\rm AdS}_5}  - \frac{p'(z_r)}{9 z_r^2} ds^2_{\Sigma_g} + \frac{kdk^2}{1-k^3} + d\varrho^2 + \varrho^2 \eta_\psi \right]\,,
\end{equation} 
and hence is regular provided that the period of $\psi$ is fixed to be $2\pi$.  The dilaton reads
\begin{equation}
e^{4\phi} \sim  \frac{1}{(F_0)^4} \frac{1}{z_r^2} \frac{(1-k^3)^3}{k^5}\,.
\end{equation}  

Similarly, near $\k = 1$ we introduce the coordinate $\varrho^2 = \frac{4}{3}(1-\k)$ and take the $\varrho \to 0$ limit. The metric takes the form
\begin{align}
ds^2_{10} &\sim  e^{2W} \left[ds^2_{{\rm AdS}_5} - \frac{\p'}{9\z^2} ds^2_{\Sigma_g} 
+ \frac{-\z p'}{3p - \z p'} \left(d\varrho^2 + \varrho^2\eta_\psi^2 \right) \right]
+ e^{-2W} d\z^2
\end{align} where the warp factor is $e^{4W} = -3 \z p/\p'$. Fixing the period of $\psi$ to be $2\pi$, the $S^1$ shrinks in a regular way. The dilaton reads
\begin{equation}
e^{4\phi} \sim \frac{1}{(F_0)^4} \frac{1}{\z^3}
\frac{(3\p)^3}{-\p'\left(3\p - \z\p'\right)^2}\,.
\end{equation}

Although in the two separate limits above the $S^1$ can shrink regularly, when the double limit is considered a singularity appears. As we will see later this is due the presence of D6-branes. 

\subsubsection{O8-plane--D8-branes}
\label{sub:O8}

In this section we study the region near $k=0$, away from $z=0$ or $p'=0$. We first make the coordinate transformation $k = f(z) r$ with
\begin{equation}
f^5(z) = -\frac{1}{z^3} \frac{3\p - \z\p'}{\p'}\,,
\end{equation}  
which maps the region near $\k=0$ to the one near $r=0$. The differential $dk$ becomes $dk =  f(z)\left(dr + r f'(z)/f(z) dz\right)$.
Following the coordinate transformation, the metric near $r=0$ takes the form
\begin{align}
ds^2_{10} &\sim r^{-1/2} ds^2_9 + r^{1/2} dr^2\,,\\
ds^2_9 &=  z^2 f^2(z) \left[ ds^2_{{\rm AdS}_5} - \frac{p'}{9z^2} \left( ds^2_{\Sigma_g} + \frac{3z dz^2}{p} + \frac{12 z^3 p}{\left(3\p - \z\p'\right)^2} \eta_\psi^2 \right)\right]\,.
\end{align}  
The dilaton reduces to 
\begin{equation}
e^{\phi} \sim \frac{1}{F_0} r^{-5/4} \,.  
\end{equation}  

The solution in this region describes a stack of $2n_8$ D8-branes, with $n_8<8$, stuck on an O8-plane. The metric and dilaton for the latter system are given by
\begin{equation}\label{eq:O8}
ds^2_{10} = H^{-1/2}_8 ds^2_{||} + H^{1/2}_8 dx_9^2 \, ,\qquad  e^{\phi} = g_s H^{-5/4}_8\, ,
\end{equation} 
where $ds^2_{||}$ is the line element of the flat spacetime parallel to the wordlvolume of the D8-branes, and $x_9$ is the transverse coordinate. $g_s$ is the string coupling constant. The function $H_8$ reads 
\begin{equation}
H_8 = c + g_s F_0 x_9\,, \qquad  F_0 = \frac{8-n_8}{2\pi \ell_s} {\rm sign}(x^9)\,,
\end{equation} 
with $F_0$ being the Romans mass,  $\ell_s$ the string length and $c$ an arbitrary constant.

The O8--D8 metric is matched to the solution near $k=0$ by the identification\footnote{The $(g_s F_0)^{-3/4}$ factor in $ds^2_{||}$ can be removed by rescaling the coordinate $z$ in $ds^2_9$.}
\begin{equation}\label{eq:c=0}
c=0\,, \qquad r = \left(g_s F_0 \right)^{1/5} x_9\,, \qquad ds^2_{||} = (g_s F_0)^{-3/4} ds^2_9\,.
\end{equation}  
While $c$ can be of any value in general, for our solution $c=0$. Since for us $F_0>0$, we have $n_8<8$. Looking at (\ref{eq:O8}), we see that this implies that the dilaton diverges on the O8-plane.  The same behavior occurs for example in the AdS$_6$ solution of massive type IIA supergravity \cite{brandhuber-oz}, which arises as the near-horizon geometry of an O8--D8--D4 system. 

In the region near $k=0$ the fluxes are non-zero, but there are no cycles that yield quantization conditions.  

As observed earlier in (\ref{eq:positivity}), $k\in [0,1]$. The presence of an O8-plane at $k=0$ suggests that our solution is, in fact, half of a bigger solution for which $k \in [-1,1]$. We can then continue our solution to $k\in [-1,0)$ by imposing the standard O8-plane conditions: $g_{\mu \nu}(-k)= g_{\mu \nu}(k)$ for $\mu,\nu\neq k$, $g_{\mu k}(-k)= -g_{\mu k}(k)$, while for the fluxes $B(-k)= -B(k)$, $F_{4i}(-k)= - F_{4i}(k)$, $F_{4i+2}(-k)= F_{4i+2}(k)$.

\subsubsection{D6-branes}
\label{sub:D6}

In this section we study the singularity in the region near $k=1$ and $p=0$.  Let $z_r$ be a (single) zero of $p$ and introduce the coordinates $(r,\theta)$ via
\begin{equation}\label{eq:rtheta}
3 r \cos^2(\theta/2) = 1 - \frac{\z}{\z_r}\,, \qquad 3r \sin^2(\theta/2) = 1-\k\,.
\end{equation}  
The region of interest maps to $r=0$ in these coordinates.  

The metric takes the form
\begin{equation}
 ds^2_{10} \sim 3z_r r^{1/2}\left[ds^2_{{\rm AdS}_5} - \frac{p'(z_r)}{9z_r^2} ds^2_{\Sigma_g} \right] + 3z_r r^{-1/2} \left[dr^2 + r^2(d\theta^2 + \sin^2(\theta) \eta_\psi^2) \right]\,. \label{D6limit}
\end{equation} 
and the dilaton 
\begin{equation}
e^{\phi} \sim \frac{1}{F_0} \frac{3^{3/2}}{z^{1/2}_r} r^{3/4}\,.  
\end{equation}  The neighborhood of $\{z=z_r,\, k=1\}$ can be identified with the region near a stack of $M$ D6-branes with ${\rm AdS_5} \times \Sigma_g$ world-volume.  The number of D6-branes $M$ is related to the parameters of the solution, as we will see shortly.  

In the string frame, the type IIA supergravity solution, that describes the geometry of a stack of $M$ D6-branes is
\begin{equation}
ds^2 = H_6^{-1/2} ds^2_{||} + H_6^{1/2} ds^2_\perp\,, \qquad e^{\phi} = g_s H_6^{-3/4} \,
\end{equation} with 
\begin{equation}
H_6 = 1+ \frac{L_6}{r}, \qquad L_6 = \frac{1}{2} M \ell_s g_s.
\end{equation} The parameters $g_s$ and $\ell_s$ are respectively the string coupling and string length.  The coordinate $r$ is the overall radial coordinate of the transverse space with metric $ds^2_\perp$.   The metric on the world-volume of the D6-branes correspond to $ds^2_{||}$.  For the solution of interest in \eqref{D6limit}, we fix the spaces as
\begin{align}
ds^2_\perp &= dr^2 + r^2(d\theta^2 + \sin^2(\theta) \eta_\psi^2)\,\\
ds^2_{||} &= L^2 \left[ds^2_{{\rm AdS}_5} - \frac{p'(z_r)}{9z_r^2} ds^2_{\Sigma_g} \right]\,
\end{align} where $L$ is the radius of the AdS$_5$.  We can identify $H_6$ from the warp factor of the transverse space and from the expression of the dilaton.  These, respectively, yield the following relations for $L_6$
\begin{equation}
L_6^{1/2} = 3 z_r\, , \qquad L_6^{1/2} = \frac{1}{9} g_s F_0\,.  
\end{equation}  These relations taken with the expression of $L_6$ in terms of $M$ yield
\begin{equation}
M = \frac{2}{3} \frac{z_r F_0}{\ell_s}\,. \label{D6N}
\end{equation}  The AdS$_5$ radius is fixed, by matching the warp factor of the world-volume space, to $L =3z_r$.  

The identification in \eqref{D6N} is made with the assumption that $r$ in \eqref{D6limit} is the radius of the space transverse to the D6-branes.  Since it was done by matching the warp factor with $H_6$ in the near-brane region, there is an overall scaling between $r$ and the radius of the transverse space that is not fixed.  However, we will compute the number of D6-branes more accurately by using the flux.

In the $\{z \to z_r,\, k \to 1\}$ limit, at leading order, the R--R fluxes read
\begin{subequations}
\begin{align}
d C_1 &\sim -\frac{1}{3} z_r F_0 \sin(\theta) d\theta \wedge \eta_\psi + \alpha_1(\theta) dr \wedge \eta_\psi\, , \\
d C_3 &\sim \frac{1}{9} \left( \kappa z_r^2 + \ell z_1^2 \right ) F_0 \sin(\theta) d\theta \wedge \eta_\psi \wedge {\rm vol}_{\Sigma_g} +  \alpha_3(\theta) dr \wedge \eta_\psi \wedge {\rm vol}_{\Sigma_g} \,  .
\end{align} 	
\end{subequations}
The explicit expressions for the functions $\alpha_1$ and $\alpha_3$ are unnecessary for the present discussion: in the region near $r=0$ the terms with a $dr$ component do not yield a quantization condition. 

The field strength $dC_1$ can be integrated over the two-sphere $S^2_{z_r}$ with coordinates $(\theta,\psi)$, and $dC_3$ can be integrated over the four-cycle $S^2_{z_r} \times \Sigma_g$.  The quantization
conditions 
\begin{equation}
\label{eq:fluxq}
\frac{1}{(2\pi\ell_s)^p} \int_{\mathcal{M}_{p+1}}  d C_p \in \mathbb{Z}
\end{equation}
for $dC_1$ and $dC_3$ respectively give 
 \begin{equation}\label{eq:Mm}
M= \frac{2}{3} \frac{z_r F_0}{\ell_s}\,,  \qquad m = \frac{V_g F_0}{18\pi^2 \ell_s^3} \left( \kappa z_r^2 + \ell z_1^2 \right) \, .
\end{equation} 
The first parameter $M$ counts the number of D6-branes and its expression is consistent with \eqref{D6N}.  The parameter $m$ counts the number of D4-branes in this region, which are dissolved in the D6-branes.  The volume $V_g$ of the Riemann surface $\Sigma_g$ is fixed by the curvature so that
\begin{equation}\label{eq:kVg}
	\kappa V_g = 4\pi (1-g)\, .
\end{equation}

There are no cycles that yield a quantization condition for the NS--NS flux.

\subsubsection{D4-branes}
\label{sub:D4}

In this section we analyze the singularities at $z=0$ and $p' =0$.  As we will see, these regions describe smeared D4-branes. In the $z=0$ case the smearing is on the Riemann surface $\Sigma_g$; in the $p'=0$ case, it is on $\Sigma_g$ and two more internal directions.  

\subsubsection*{$z\to 0$ limit}

In the region near $z=0$ we make the coordinate transformation $\{ z^3= (\frac{3}{2}r)^{2}$, $k^3=\cos^{2}(\theta) \}$,
and write the metric as
\begin{align}
ds^2_{10} &\sim  \left(\frac{3}{2} r \cos(\theta)\right)^{-1/3} 
\left[Q^{-1/2} r^{2/3} ds^2_{{\rm AdS}_5} + Q^{1/2} r^{-2/3} ds^2_5 \right] \,, \nn \\
ds^2_5 &=  \frac{1}{3} \p(0) ds^2_{\Sigma_g} 
+ dr^2 + r^2\left(d\theta^2 + \sin^2(\theta) \eta_\psi^2 \right) \, ,
\end{align}
where $Q \equiv \left(\frac{3}{2}\right)^{-4/3} \ell \z^2_1/\p(0)$. The dilaton reads:
\begin{equation}
e^\phi \sim \frac{1}{F_0} Q^{-1/4} r^{1/3} \left(\frac{3}{2} r \cos(\theta)\right)^{-5/6}\, .
\end{equation}

This metric describes a stack of D4-branes with AdS$_5$ world-volume, which are smeared on the Riemann surface, $\Sigma_g$, while being inside a stack of D8-branes. We describe this system in more detail in appendix \ref{app:D4inD8}. To identify the number of the D4-branes, we look at the $dC_3$ field strength in this region:
\begin{align}
dC_3 \sim - \frac{4}{9} \ell z_1^2 F_0\, k dk \wedge \eta_\psi \wedge  {\rm vol}_{\Sigma_g} + \beta_3(k) dz \wedge \eta_\psi \wedge  {\rm vol}_{\Sigma_g}  
\end{align}
The quantization condition for $dC_3$ then yields
\begin{equation}\label{eq:nD4}
n = \frac{V_g F_0}{18\pi^2 \ell_s^3} \ell z_1^2\, .
\end{equation} 
$n$ is an integer: its absolute value the number of D4-branes in this region. 

The R--R field strength $dC_1$ vanishes in this region, and there are no cycles that can provide a quantization condition for the NS--NS flux. However, we point out that the latter is singular near $z=0$:
\begin{align}
H \sim \frac{z_1^2}{3z} \left( dk - \frac{dz}{z} \right) \wedge  {\rm vol}_{\Sigma_g}\,. 
\end{align}

The singularity of $H$ in this region is a phenomenon which we do not fully understand; an analogous singularity in the region near the D6-branes is discussed in section \ref{sub:flux}.

\subsubsection*{$p' \to 0$ limit}

We now turn to the $p'= 0$ region. The function $p'$ can vanish at $z=z_1$ for $\kappa=\ell$.  In this region, we introduce the coordinate $r = \frac{3}{2}\kappa(1-\z/\z_1)$. In the limit $r\to 0$ and upon making the coordinate transformation $\k=\cos^{2/3}(\theta)$, the metric takes the form:
\begin{align}
ds^2_{10} &\sim  \left(\cos(\theta)\right)^{-1/3}
\left[(\tilde{Q} r)^{-1/2} ds^2_{{\rm AdS}_5} + (\tilde{Q} r)^{1/2} ds^2_5 \right]\,, \nn \\
ds^2_5 &= \frac{4}{9} \tilde{Q}^{-1} ds^2_{\Sigma_g} 
+  \frac{4}{9}z^2_1\left(dr^2 + d\theta^2 + \sin^2(\theta) \eta_\psi^2 \right)\,.
\end{align}  
where $\tilde{Q} \equiv \frac{4}{3} \z_1/\p(\z_1)$. The dilaton reads:
\begin{equation}
e^{\phi} \sim \frac{1}{F_0} \frac{1}{z_1} (\tilde{Q} r)^{-1/4} \left(\cos(\theta)\right)^{-5/6}\,.
\end{equation}

In this case the local behavior of the metric can be recognized as an example of the ``harmonic function rule'' for delocalized branes \cite{Papadopoulos:1996uq,Tseytlin:1996bh,Gauntlett:1996pb}. It describes a stack of D4-branes with AdS$_5$ world-volume, smeared on the Riemann surface $\Sigma_g$ and on the two-sphere with coordinates $(\theta,\psi)$. The overall warp factor blows up at $\theta= \pi/2$, and this due to the fact that there is a O8--D8 system localized there. 

The $dC_3$ field strength in this region reduces to 
\begin{align}
dC_3 \sim - \frac{4}{9} \ell z_1^2 F_0 \, k dk \wedge \eta_\psi \wedge  {\rm vol}_{\Sigma_g}  + \gamma_3(k) dz \wedge \eta_\psi \wedge  {\rm vol}_{\Sigma_g}\,,
\end{align} 
and yields the following quantization condition:
\begin{align}\label{eq:nsmD4}
n = \frac{V_g F_0}{18 \pi^2 \ell^3_s} \ell z_1^2\,, 
\end{align} 
This coincides with \eqref{eq:nD4}, the integer $n$ counting the number of D4-branes. 
There are no cycles that can provide a quantization condition for the R--R field strength $dC_1$ and the NS--NS flux in this region.

\subsection{The range of the coordinates}
\label{sub:range}

In this section, we determine the possible ranges for the coordinates $\psi$, $\k$ and $\z$.  Each choice of ranges will correspond to different classes of solutions.  

The regularity analysis in the previous sections restricts the circle coordinate $\psi$ and the coordinate $k$ as
\begin{equation}
\psi \in [0,2\pi] \, ,\qquad k \in [0,1]\, .  
\end{equation}

There are more options for $\z$ since there are three ``special'' regions: $0$, $z_r$ and $z_1$ where there are brane sources. The possible intervals depend on the ordering of these points along $z$ and on the positivity conditions \eqref{eq:positivity} for a given choice of $\kappa$ and $\ell$.  In what follows we use these conditions to identify the possible choices; we will summarize the result at the end.  

First consider the cases when $\kappa = 0$ or when $\kappa = -\ell$.  The only real root of $\p$ is $\z_0$, whereas $\p'$ has no real roots.  The positivity condition on $p'$ implies that $\kappa =-\ell =-1$.  For these cases $z \in [0,z_0]$, and obviously $z_0>0$.

When $\kappa = \ell$, $\p'$ has roots at $\z = \pm \z_1$ and $\p$ can have three real roots. The system is analyzed by writing $\p$ as
\begin{equation}
\p = \kappa (\z-z_0)(\z-z_-)(\z-z_+)\, ,
\end{equation}
where the roots satisfy
\begin{equation}\label{eq:rootscond}
z_0 + z_- + z_+ = 0\,, \qquad 3 \z_1^2 = - z_0 z_- - z_- z_+ - z_+ z_0\,.
\end{equation} Without loss of generality we can take $z_0$ to be always real, whereas 
\begin{equation}
z_\pm \in \mathbb{R} \qquad \mbox{only when} \qquad z_0^2 \leq 4 z_1^2.
\end{equation}
When $z_\pm$ are imaginary, the positivity conditions on $p$ and $p'$ yield 
\begin{equation}
z \in  \begin{cases} 
\ [0,z_1]  &\mbox{for} \quad \{\kappa = +1,\ z_0 < -2 z_1\}\,, \vspace{.1cm} \\ 
\ [z_1, z_0]  &\mbox{for} \quad \{\kappa =-1,\ z_0 > 2 z_1\}\,. \end{cases}  
\end{equation}

When the roots are all real, there is a permutation symmetry among them; moreover at least one is negative and at least one is positive.  Without loss of generality $\z_-$ can be chosen to be the smallest negative root and $\z_0$ the smallest positive root.  Depending on whether $\z_+$ is negative or positive there are two orderings of the roots of $\p$ and $\p'$: (a) $(\z_-, -\z_1, \z_+, 0 , \z_1, \z_0)$ and (b) $(\z_-, -\z_1, 0 , \z_0, \z_1, \z_+)$.  The positivity conditions yield
\begin{equation}
z \in  \begin{cases} 
\ [0,z_0]  &\mbox{for} \quad \{\kappa =+1,\ 0 < z_0 \leq z_1\} \, ,\vspace{.1cm} \\ 
\ [z_1, z_0] &\mbox{for} \quad \{\kappa =-1,\ z_1 < z_0 \leq 2z_1\}\, . \end{cases}  
\end{equation}
\

The possible choices of intervals for $z$ consistent with the positivity conditions in \eqref{eq:positivity}  can be summarized as follows:\footnote{Recall that we are taking $z\geq 0$ and $z_1\geq 0$ without loss of generality.}
\begin{enumerate}
\item $\z \in [0,\z_0]$ \\
This interval is realised for $\ell = 1$ and all values of $\kappa$, with the restriction $z_0 \leq z_1 $ for $\kappa = 1$ and $z_1 > 0$ for $\kappa = 0$.

\item $\z \in [0,\z_1]$ \\
This interval is realised for $\ell = \kappa = 1$, with the restriction $z_0 < -2 z_1$ and $z_1>0$.

\item $\z \in [\z_1, \z_0]$ \\
This interval is realised for $\ell = \kappa = -1$, with the restriction $z_1 < z_0$.
\end{enumerate}  

Each of these cases correspond to a different branch of the space of solutions, and describes a different class of theories.  We will explore these three cases in detail in section \ref{sec:cases}.

\subsection{NS--NS Flux quantization} 
\label{sub:flux}

In the discussion of the various sources in the main solution \eqref{fullsol}, the possible quantized fluxes from the R--R potentials were computed and identified with the charges of various branes.  There is an additional quantization condition that comes from integrating the NS--NS flux $H$ over the whole internal space $M_3$: it measures the number of NS5-branes wrapping the Riemann surface, whose near-horizon limit yields the AdS$_5$ solution.  In this section we discuss this quantization condition, carefully considering some of the regularity issues regarding the NS--NS potential $B$ given in \eqref{eq:B}.  

The NS--NS flux is $H=d B$ on the domain $D$ defined by the range of the coordinates $k$ and $z$. While $k\in [0,1]$, there are three possible ranges for $z$, as discussed in the previous section. Let us denote a given range of $z$ as $ [z_{\rm L}, z_{\rm R}]$ and write the domain as $D = [0,1]\times [z_{\rm L},z_{\rm R}]$. By Stokes' theorem, the flux integral is $\int_{D\times S^1} H = \int_{\del D \times S^1} B$ where the $S^1$ is the $\psi$-circle. From the component under consideration in \eqref{eq:B} i.e.\ the $dk\wedge \eta_\psi$ term, we obtain
\begin{equation}\label{eq:intH1}
	\int H = -\frac43 \pi \left[\frac{z^2 p'}{3p-z p'}\right]_{z_{\rm L}}^{z_{\rm R}}\, .
\end{equation}
The relevant values of the function in brackets are
\begin{equation}
	\frac{z^2 p'}{3p-z p'} = \begin{cases}
		0 \quad &\mbox{for} \qquad  z = 0 \, ,\\
		0  \quad &\mbox{for} \qquad z = z_1  \quad \mbox{and} \quad \kappa = \ell \,, \\
		-z_r \quad &\mbox{for} \qquad z = z_r  \quad \mbox{and} \quad p(z_r) =0\,.
	\end{cases}
\end{equation}
From the summary in section \ref{sub:range}, we see that there is nonzero flux only for cases 1 and 3, given as
\begin{equation}\label{eq:intH}
	\int H = \frac43\pi z_r \, .
\end{equation}
Flux quantization requires 
\begin{equation}\label{eq:fluxH}
	\frac1{4\pi^2\ell_s^2}\int H \equiv N \in \zz  \, .
\end{equation}

In the quantization of the NS--NS flux, there are no subtleties regarding the regularity of $B$ at the boundary of $D$.  This is not the case in the quantization of the R--R flux. For example, the gauge choice for $B$ could influence the result for the number of D6-branes computed in section \ref{sub:D6}. So we need to worry about regularity near that region. First, let us analyze the $k\to 1$ and $z\to z_r$ limits separately. 

The standard way to recognize a regular form is to transform to local Cartesian coordinates. If $r$ is the local radial coordinate where a circle shrinks, so that the metric contains a $dr^2 + r^2 d\psi^2$ ``piece'', we can transform to local Cartesian coordinates $x= r \cos \psi$,  $y = r \sin \psi$. Then we see for example that $r^2 d \psi= xdy-ydx$ and $r dr \wedge d \psi= dx\wedge dy$ are regular forms, while $d \psi=\frac{xdy-ydx}{x^2+y^2}$ is not. 

Around $k=1$, the radial coordinate is $r = \sqrt{1-k}$ and $B \sim r dr \wedge d \psi + (1-r^2) {\rm vol}_{\Sigma_g}$, which is regular (the dependence on $z$ is suppressed). On the other hand, around $z=z_r$ the radial coordinate is $r = \sqrt{z_r-z}$ and $B$ has a $dk \wedge d\psi$ component which is not multiplied by a vanishing function of $r$; this is not regular. 

Thus we have to find another gauge for $B$. A simple choice (but by no means the only one) is to perform the gauge transformation 
\begin{equation}\label{eq:B'}
	B' = B + d \Lambda_1 \, ,\qquad \Lambda_1 =  \frac1{F_0}\left(1-\frac 1k\right)C_1\, ,
\end{equation}
where $C_1$ was given in (\ref{eq:C1}). This choice does not spoil regularity of $B$ at $k=1$, while at $z=z_r$ it cures the regularity problem we just saw: the coefficient of $d k\wedge d \psi$ now has a double zero in $r$.

We now have to take care of the simultaneous limit $\{z\to z_r,\, k\to 1\}$ which, as we saw in section \ref{sub:D6}, yields a region where  D6-branes are located. At that locus we cannot expect $B$ to be regular and in fact this is not the case even for $H$, as near that region it has a $r dr \wedge {\rm vol}_{S^2}\sim r^{-1} {\rm vol}_{{\Bbb R}^3}$ component. This is the same behavior that the NS--NS flux of the AdS$_7$ solutions of \cite{afrt} exhibits near a similar locus. In particular, this is the case for the solution one obtains by reducing the AdS$_7\times S^4/{\Bbb Z}_k$ solution of M-theory to IIA. As can be seen from \cite[Eq.~(5.7)]{afrt}, the radial coordinate there being $r=\alpha^2$, $H\sim r d r \wedge {\rm vol}_{S^2}\sim r^{-1} {\rm vol}_{{\Bbb R}^3}$. While this might look puzzling at first, it comes about as one reduces a regular 4-form flux in M-theory along a shrinking $S^1$.  

What we want to impose on $B$ then, is that its exterior derivative does not contain any delta functions.\footnote{This criterion was also imposed for the aforementioned AdS$_7$ solutions.} This can be checked by performing an integral of $B$ over a path $\gamma$ that goes from $\{k=1\}$ to $\{z=z_r\}$, and taking the limit of the result $\int_\gamma B$ as $\gamma$ shrinks to the point $\{z=z_r,\, k=1\}$. If $\int_\gamma B\to 0$, then $dB$ does not contain any delta functions. For the gauge (\ref{eq:B'}), this is indeed the case.

Although R--R flux quantization, which we carried out in section \ref{sub:D6} to count the number and charges of D6-branes, would be  more strictly performed in the gauge (\ref{eq:B'}), the result is eventually what we obtained in that section, namely (\ref{eq:Mm}). This is because the gauge transformation $\Lambda_1$ in (\ref{eq:B'}) goes to zero at $\{z=z_r,\, k=1\}$.
In any case, the result for the number of D6-branes could also be obtained more simply from the Bianchi identity $dF_2 - H F_0 = \delta$. By integrating this over the entire $M_3$ and using (\ref{eq:fluxq}) and (\ref{eq:fluxH}), we obtain 
\begin{equation}\label{eq:NnM}
	N n_0 = M\, .
\end{equation}
The above equation, combined with (\ref{eq:intH}) indeed gives (\ref{D6N}).


\subsection{Central charge} 
\label{sub:a}

A handle on a CFT$_4$ is provided by the central charges which given a dual AdS$_5$ solution can be computed holographically, following \cite{henningson-skenderis}. In particular, the central charge $a$ is related to the five-dimensional Newton constant $G_5$ via 
\begin{equation}
a = \frac{\pi R_{{\rm AdS}_5}^3}{8G_5}\, .
\end{equation} 
$G_5$ can be obtained by compactifying the ten-dimensional action. In our case, $R_{{\rm AdS}_5}$ in fact depends on the internal coordinates, since we are dealing with a warped compactification. A reasonable procedure (as in \cite{gaiotto-t-6d,cremonesi-t}, for example) is to average the warping function coming from $R_{{\rm AdS}_5}^3$ over $M_5$. This leads to
\begin{equation}
\frac{1}{G_5} = \frac{1}{G_{10}} \int e^{8W-2\phi} {\rm vol}_5\, , \qquad 16\pi G_{10} = \frac{\left(2\pi \ell_s\right)^8}{2\pi}\, ,
\end{equation} 
where ${\rm vol}_5$ is the volume form of the internal manifold, defined by $(-p'/9z^2)ds^2_5$ in \eqref{fullsol}. The factor $e^{-2\phi}$ appears as we switch from the string to the Einstein frame.   We find
\begin{equation}
e^{8W-2\phi}{\rm vol}_5 = \frac{2}{27} F_0^2 (-z^2 \p') dz \wedge k dk \wedge d\psi \wedge {\rm vol}_{\Sigma_g}\, .
\end{equation}
The central charge is then
\begin{equation}\label{eq:a-gen}
a = \frac{8}{27} \frac{\pi^4 V_g}{(2\pi\ell_s)^8} F_0^2 \int (-z^2\p') dz = \frac{8}{27} \frac{\pi^4 V_g}{(2\pi\ell_s)^8} F_0^2 \left[\ell z_1^2 z^3-\frac35 \kappa z^5 \right]_{z_{\rm L}}^{z_{\rm R}}\, .
\end{equation}
In the next section we will evaluate $a$ for the three different classes identified at the end of section \ref{sub:range}.


\section{Three classes of solutions} 
\label{sec:cases}

At the end of section \ref{sub:range}, we concluded that there are three possibilities for the range of the coordinate $z$. These actually correspond to different physics. We will now study them in turn. 

\subsection{O8, D6, D4}
\label{sub:6d}

\begin{figure}[ht]
\centering	
	\includegraphics[width=8cm]{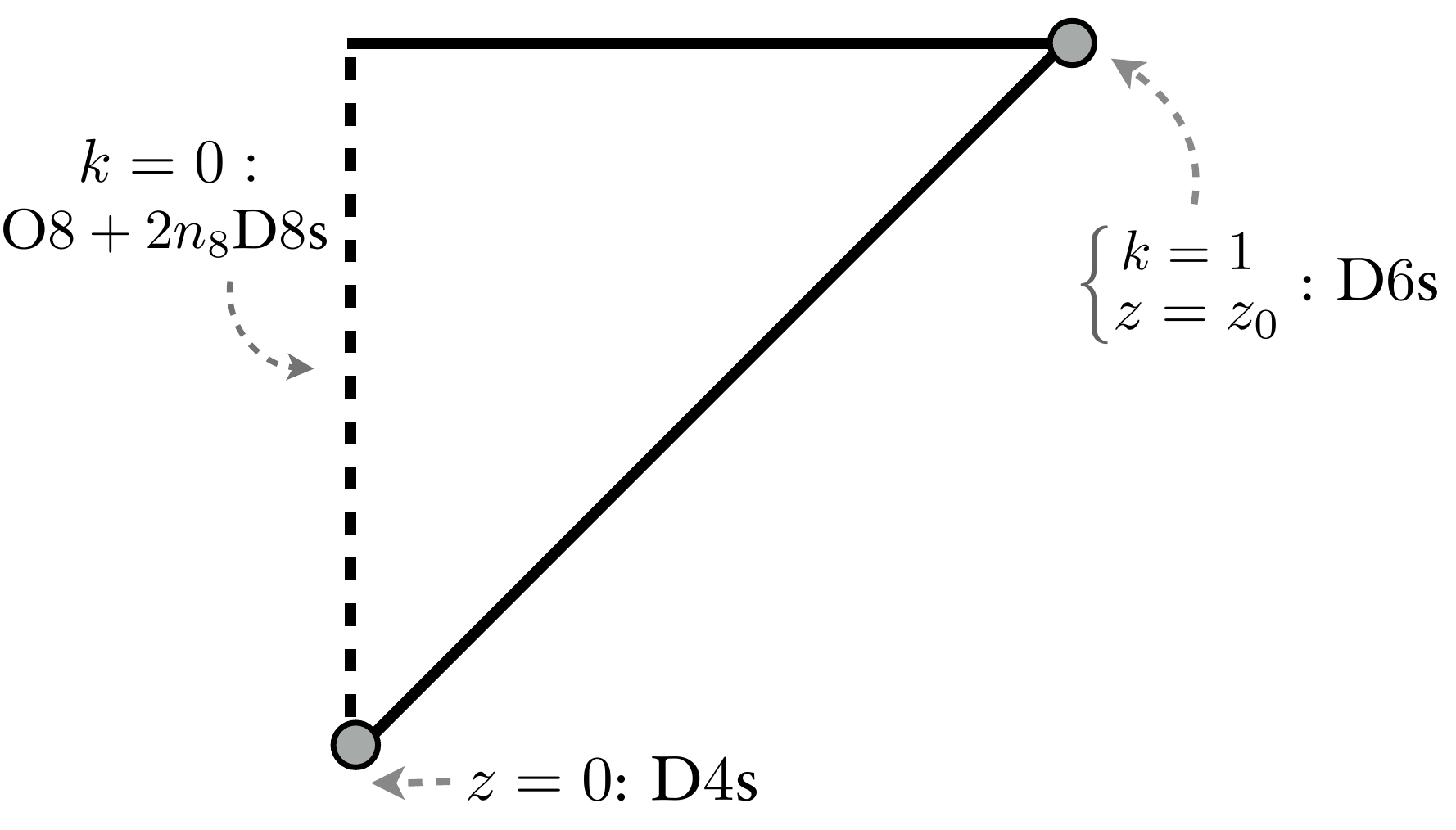}
	\caption{\small A cartoon of the internal space $M_3$ corresponding to the range $z\in [0,z_0]$. Roughly, $z$ runs vertically and $k$ horizontally. At $\{z=z_0,\, k=1\}$ the solution has a D6-brane stack. There is an O8-plane (with $n_8$ D8-brane pairs on top) at $k=0$ and $z=0$. In fact, the locus $z=0$ is revealed to be a ``blowup'' of a D4-brane stack inside an O8-plane, in the sense that near that locus, $k$ becomes one of the angular coordinates of the sphere transverse to the D4-branes. Finally, at $z=z_0$ the metric is regular.}
	\label{fig:0z0}
\end{figure}

We start from the case $z\in [0,z_0]$ and $\ell=1$. Following the analysis of sections \ref{sub:O8}, \ref{sub:D6} and \ref{sub:D4}, there are 
\begin{itemize}
	\item an O8--D8 stack at $k=0$; 
	\item D6-branes at $\{z=z_0,\, k=1\}$; 
	\item D4-branes at $z=0$, smeared on the Riemann surface $\Sigma_g$. 
\end{itemize}
The number $M$ of D6-branes and $n$ of D4-branes are related to the parameters of the solution by (\ref{eq:Mm}) and (\ref{eq:nD4}), which we repeat here for the reader's convenience: 
\begin{equation}\label{eq:Mn}
		M = \frac{n_0}{3\pi \ell_s^2} z_0\, , \qquad n = \frac{n_0 V_g}{36 \pi^3 \ell_s^4} z^2_1\, .
\end{equation}
The flux quantum $n_0 = 2\pi \ell_s F_0$ is related by $n_0 = 8-n_8$ to the number $n_8$ of D8-brane pairs on top of the O8-plane. From (\ref{eq:Mm}) and (\ref{eq:kVg}) we also see that the D6-branes have a D4-brane charge $m= n + (1-g) M^2/n_0$, or 
\begin{equation}
	m=n + (1-g)N M
\end{equation}
in terms of the NS--NS flux quantum $N$; see (\ref{eq:NnM}). This is manifestly an integer. The situation is summarized in figure \ref{fig:0z0}.

Because of the rescaling (\ref{eq:resc}), the space of solutions can be parameterized by $z_1/z_0$. From (\ref{eq:Mn}) and(\ref{eq:NnM}) it follows
\begin{equation}\label{eq:z1z0}
	\frac{z_1^2}{z_0^2}= \frac{4\pi}{V_g}\frac{n}{MN} \, .
\end{equation}
Hence, the space of solutions is discretized by flux quantization.

By specializing (\ref{eq:a-gen}) to the $[0,z_0]$ interval, we find 
\begin{equation}\label{eq:a-0z0}
	a=\frac{27}{32}\left(\frac15 (g-1) N^3 M^2 + \frac13 n N^2 M \right)
\end{equation}
as anticipated in (\ref{eq:a}). The $N^3 M^2$ term points to a compactification of a six-dimensional field theory. We will soon make this more precise, and identify the theory. $n$ is the number of D4-branes, and it is natural to interpret them as punctures, similar to the interpretation of M5-branes wrapping the R-symmetry circle as punctures, in \cite{gaiotto-maldacena,Bah:2013wda}. Thus, the $n N^2 M$ term is the contribution of the punctures to the central charge. Since the D4-branes are uniformally distributed on the Riemann surface, we interpret these as simple punctures. 

Let us now have an additional look at the various cases for the Riemann surface $\Sigma_g$. 

\subsubsection{Genus greater than one}
\label{sub:kappa-1}

For $\kappa = -1$ there are no constraints on $z_1$ and $z_0$, other than $z_0$ being positive ($z_1$ was fixed to be non-negative, without loss of generality, in section \ref{sec:new_solutions}.) As mentioned before, the space of solutions can be parameterized by the ratio $z_1/z_0$. In this case it is a half-line. 

It is worth investigating what happens at the extremum of this half-line, namely for $z_1=0$. Given (\ref{eq:z1z0}), it corresponds to $n=0$, a solution without punctures. The change of coordinates 
\begin{equation}
	\frac{z^3}{z_0^3}=1- (1-w^3) \cos^2(\theta/2) \,,\qquad \frac{z}{z_0} k = w
\end{equation}
transforms the metric of the solution to 
\begin{equation}\label{eq:ads5nop}
	ds^2_{10} =  e^{2W} \left[ds^2_{{\rm AdS}_5} + \frac{1}{3} ds^2(\Sigma_g) + \frac{wdw^2}{1-w^3} + \frac{1-w^3}{9} \left(d\theta^2 + \sin^2(\theta) \eta_\psi^2 \right) \right]\,.
\end{equation} The warp factor and dilaton read
\begin{equation}
e^{4W} = z_0^2 \frac{1-w^3}{w}\,, \qquad e^{\phi} = \frac{z_0}{F_0}\left(\frac{1-w^3}{w} \right)^{5/4}\,.
\end{equation}
This solution is a member of the family of solutions obtained in \cite{afpt},\footnote{It corresponds to the value $18$ of the parameter $b_2$ that parameterizes the solutions in \cite{afpt}. The coordinate $w$ is related via $w^2 = y/y_0$ to the coordinate $y$ appearing there, where $2y_0 = F_0 z_0^2$.} which, as argued there, are dual to compactifications of six-dimensional field theories on Riemann surfaces $\Sigma_{g>1}$ of negative curvature, without punctures.\footnote{The central charge for similar theories was computed in \cite{10letter} and agrees with (\ref{eq:a-0z0}), although the factor $27/32$ was mistakenly omitted there.} This argument was later strengthened in \cite{prt}, where the holographic RG flow connecting the AdS$_7$ duals of the six-dimensional field theories and the AdS$_5\times \Sigma_{g>1}$ solutions was obtained.  

This limiting case supports our earlier claim that the solutions in this class should be interpreted as dual to compactifications of a six-dimensional field theory. As we have just seen, they generalize the AdS$_5$ solutions of \cite{afpt} by including D4-branes smeared over $\Sigma_g$. In \cite{afpt}, where no punctures were present, only the case of genus $g>1$ was realised. With the inclusion of punctures, we are able to obtain also $g=0$ and $g=1$. In section \ref{sec:6d} we will discuss the aforementioned six-dimensional field theory and its AdS$_7$ gravity dual.

\subsubsection{Genus one}
\label{sub:g1}
For $\Sigma_{g=1}=T^2$ ($\kappa = 0$), the only constraint is $z_1>0$. This matches with the fact that in this case there should be no solutions without punctures ($n=0$ or $z_1 =0$), as found in \cite{afpt}. Since $p(z)$ in \eqref{p(z)} becomes linear, the solution is particularly simple. Moreover, the directions of $T^2$ are isometries. 

One can T-dualize to type IIB supergravity along one of these isometries. The dilaton is not constant on the IIB side; in particular the T-dual solution is not a Sasaki--Einstein compactification. 

From (\ref{eq:a-0z0}) we see that the $N^3M^2$ term drops out of the central charge. This is similar to what happens for the class $\mathcal{S}$ theories, where the compactification on a torus results in circular quiver of ordinary gauge theories, with $a \sim nN^2$ growth.

\subsubsection{Genus zero}
\label{sub:g0}

For $\Sigma_{g=0}=S^2$ ($\kappa=1$), we saw in \ref{sub:range} that $z_1 \ge z_0$. Via (\ref{eq:z1z0}) and (\ref{eq:kVg}) this constraint translates to the bound
\begin{equation}\label{eq:bound}
	n\ge MN \, .
\end{equation}
In other words, there is a minimum amount of punctures one can have on the sphere. This again has a counterpart in the class $\mathcal{S}$ theories, where for example one cannot have three simple punctures on a sphere. 

It is natural to investigate what happens when the bound is saturated. We analyze this in appendix \ref{app:z1=z0}. Unfortunately, the physical interpretation of this limiting case is challenging, and it is not clear what it represents.

\subsection{O8, D4 distribution} 
\label{sub:5d}

The second case we found in section \ref{sub:range} is $z\in[0,z_1]$. As explained there, this is only possible for $\ell=\kappa=1$. We can again parameterize the solutions by the ratio $z_1/z_0$, which has to lie in the range $(-\frac12,0)$.

\begin{figure}[ht]
\centering	
	\includegraphics[width=7cm]{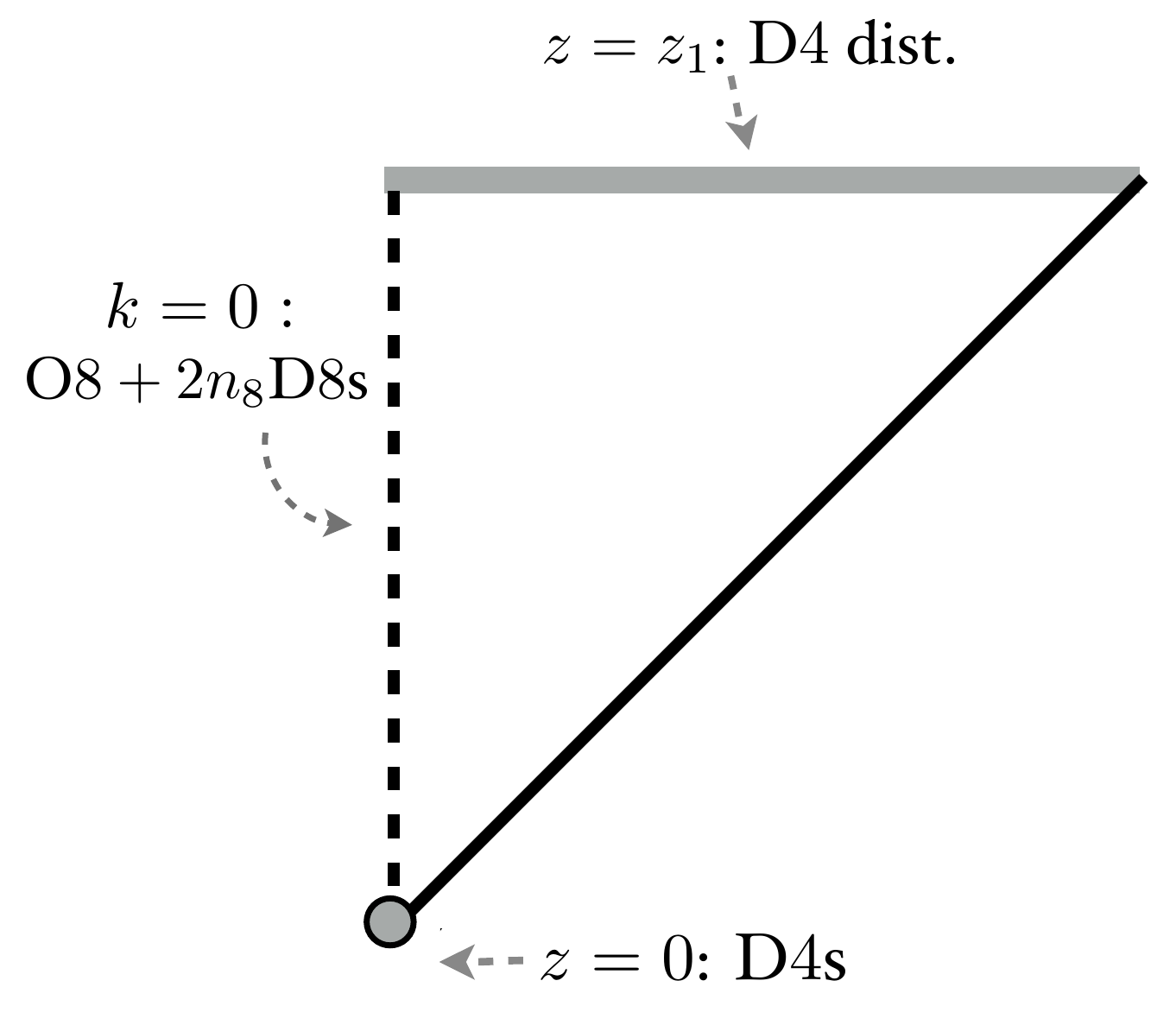}
	\caption{\small A cartoon of $M_3$ for the range $z\in [0,z_1]$. Compared to the solution in figure \ref{fig:0z0}, the solution has no D6-branes; the upper range for $z$ is at $z=z_1$, where a distribution of D4-branes appears.}
	\label{fig:0z1}
\end{figure}

In this case, $z$ does not reach the $z=z_0$ locus; as a result, there are no D6-branes in the solution. On the other hand, at $z=z_1$ there is a uniform distribution of D4-branes smeared not only on $\Sigma_{g=0} = S^2$ (as was the case for the solutions of section \ref{sub:6d}) but also on the ($k,\psi$) directions of the internal $M_3$. The total number of the D4-branes is equal to 
\begin{equation}\label{eq:nz12}
	n= \frac{n_0}{9\pi^2 \ell_s^4} z^2_1\, ,
\end{equation}
which is the same total number as that of the D4-branes at $z=0$.  The situation is depicted by figure \ref{fig:0z1}. 

The absence of D6-branes ($M=0$) is also reflected in the vanishing of the NS--NS flux quantum, $N=0$. This can be seen from (\ref{eq:NnM}), or directly from (\ref{eq:intH1}). Using (\ref{eq:a-gen}) we evaluate the central charge: 
\begin{equation}
	a= \frac 9{5\cdot 16}\frac{n^{5/2}}{n_0^{1/2}}\, .
\end{equation}
This $n^{5/2}$ scaling is the case also for the AdS$_6$ solution of \cite{brandhuber-oz}, which similar to the present setup involves $n$ D4-branes on top of an O8--D8 system. (This scaling was also reproduced in field theory \cite{jafferis-pufu}.) This seems to suggest that the CFT$_4$ dual to the AdS$_5$ solution under consideration, has a five-dimensional origin.

The limiting case $z_1/z_0=-1/2$ is the same as the one discussed in appendix \ref{app:z1=z0}. 

\subsection{O8, D6, D4 distribution}

Finally we describe the case $z\in[z_1,z_0]$. As we saw in section \ref{sub:range}, this is only possible for $\ell=\kappa=-1$. The ratio $z_1/z_0$ has to lie in the range $[0,1)$.

As in the class of section \ref{sub:6d}, there are D6-branes at $\{z=z_0,\, k=1\}$. However, at $z=z_1$ there is a uniform distribution of D4-branes similar to the one of section \ref{sub:5d}. Again, the number of D6-branes and D4-branes are given by (\ref{eq:Mm}) and (\ref{eq:nD4}): 
\begin{equation}
		M= \frac{n_0}{3\pi \ell_s^2} z_0\,, \qquad n = \frac{n_0(1-g)}{9\pi^2 \ell_s^4} z^2_1\, .
\end{equation}

\begin{figure}[ht]
\centering	
	\includegraphics[width=8cm]{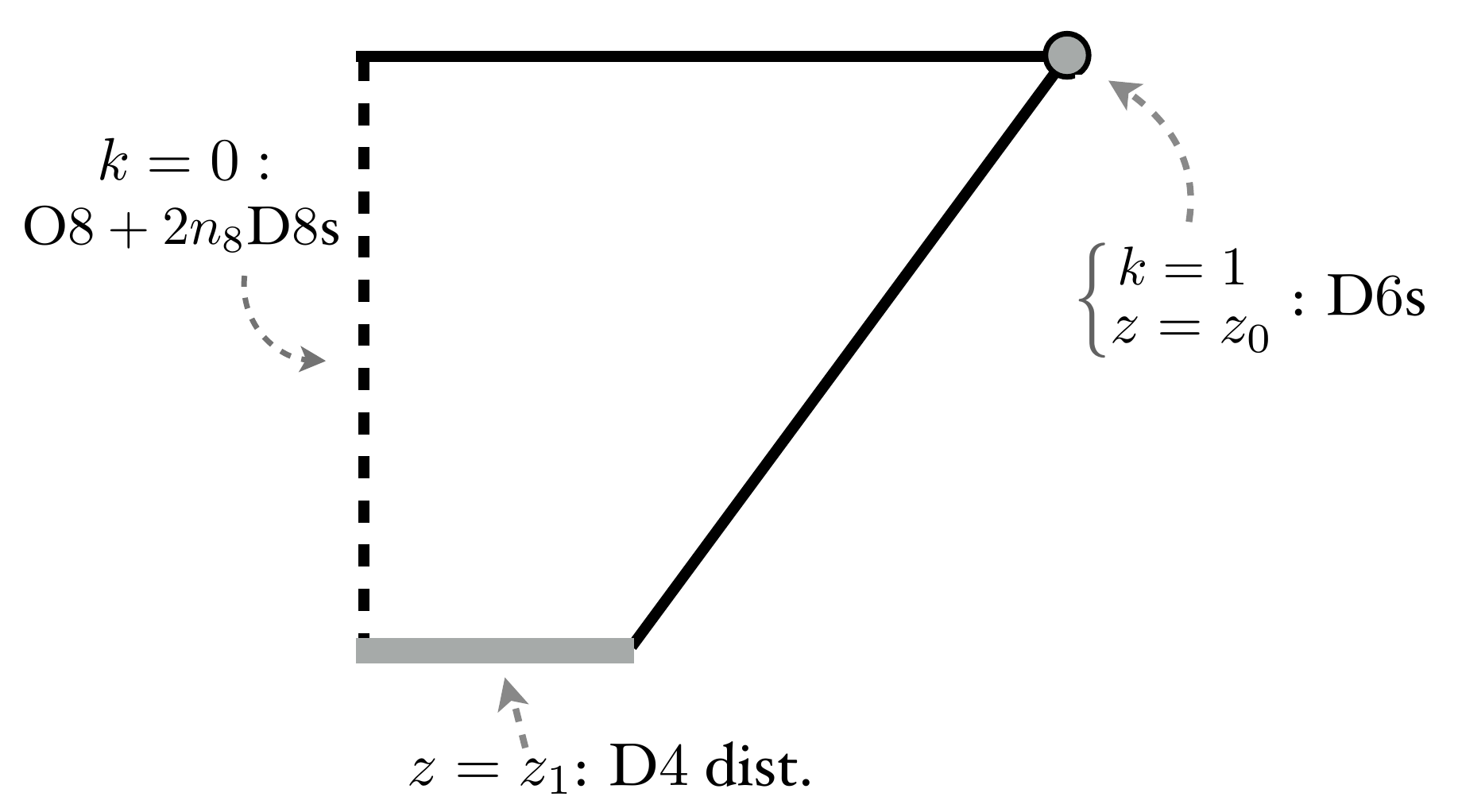}
	\caption{\small A cartoon of $M_3$ for the range $z\in [z_1,z_0]$. Compared to the solution in figure \ref{fig:0z0}, the lower range for $z$ is at $z=z_1$, where a distribution of D4-branes appears; in other words, the D4-branes in figure \ref{fig:0z0} have spread out in the $k, \psi$ directions as well.}
	\label{fig:z1z0}
\end{figure}

The central charge is given this time by 
\begin{equation}
	a= \frac{27}{32}\left(\frac15 N^3 M^2 (1-g)+ \frac13 nN^2M + \frac2{15} \frac{n^{5/2}}{n_0^{1/2}(1-g)^{3/2}}\right)
\end{equation}
which seems to signal a mix of the six-dimensional origin of section \ref{sub:6d} and of the conjectural five-dimensional origin of section \ref{sub:5d}. 

The limiting case $z_1=0$ turns out to be again the punctureless solution with $\kappa=-1$ discussed in section \ref{sub:kappa-1}, a member of the family of solutions of \cite{afpt}. 

\medskip 

\begin{figure}[ht]
\centering	
	\includegraphics[width=17cm]{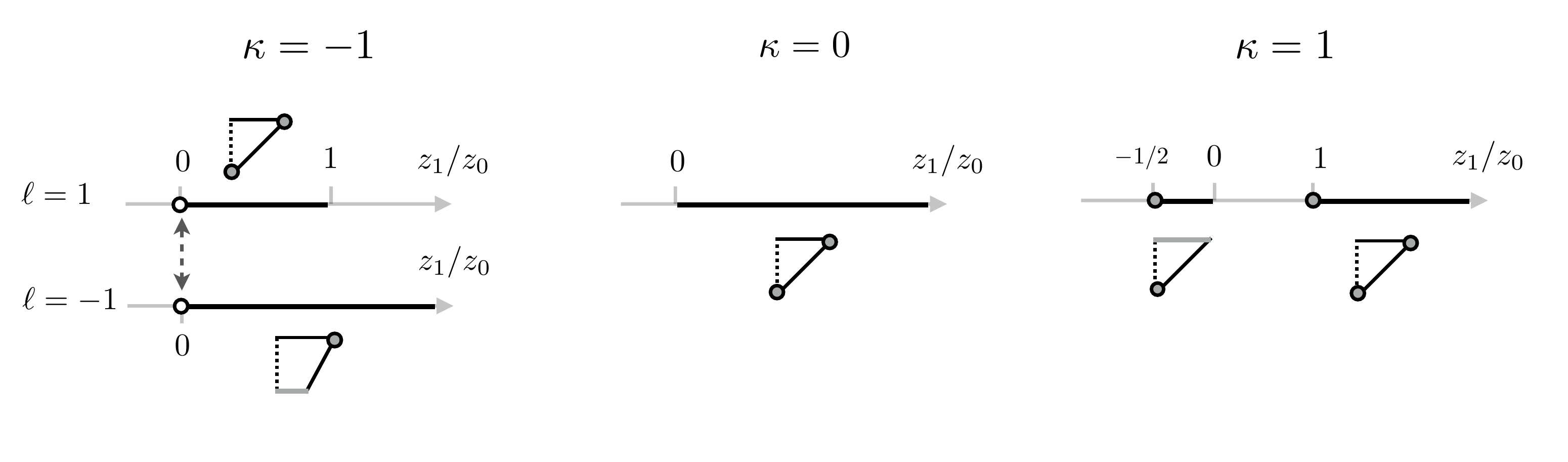}
	\caption{\small A summary of the various solutions we found for $\kappa=0,-1,1$. The cartoons refer to figures \ref{fig:0z0}, \ref{fig:0z1} and \ref{fig:z1z0}. The grey dot for $\kappa=1$ is the solution discussed in appendix \ref{app:z1=z0}. The white dots for $\kappa=-1$ both represent the punctureless solution of \cite{afpt}, whose six-dimensional origin we discuss in section \ref{sec:6d}.}
	\label{fig:summary}
\end{figure}

All the various cases we discussed are summarized in figure \ref{fig:summary} in terms of the parameter $z_1/z_0$. 

\section{The six-dimensional theory and its gravity dual}
\label{sec:6d}

In this section we will describe the six-dimensional origin of the solutions of section \ref{sub:6d}. We will first describe the AdS$_7$ solutions (which are a particular case of \cite{afpt}, but which were not appropriately discussed there), and then their CFT$_6$ duals.

\subsection{The ${\rm AdS}_7$ solution} 
\label{sub:ads7}
 
Let us first review the family of AdS$_7$ solutions of type IIA supergravity \cite{afrt, 10letter}. We will present the solutions as in \cite{cremonesi-t}\footnote{We will call here $\zeta$ the coordinate that in \cite{cremonesi-t} was called $z$.}. The metric, fluxes and dilaton read\footnote{$B$ is to be understood up to a gauge transformation.}
\begin{subequations}\label{eq:ads7a}
	\begin{align}
		&\frac1{\pi \sqrt2} ds^2= 8\sqrt{-\frac \alpha{\ddot \alpha}}ds^2_{{\rm AdS}_7}+ \sqrt{-\frac {\ddot \alpha}\alpha} \left(d\zeta^2 + \frac{\alpha^2}{\dot \alpha^2 - 2 \alpha \ddot \alpha} ds^2_{S^2}\right)\, ;\\
		&B=\pi \left( -\zeta+\frac{\alpha \dot \alpha}{\dot \alpha^2-2 \alpha \ddot \alpha}\right) {\rm vol}_{S^2}\ ,\qquad F_2 =  \left(\frac{\ddot \alpha}{162 \pi^2}+ \frac{\pi F_0\alpha \dot \alpha}{\dot \alpha^2-2 \alpha \ddot \alpha}\right) {\rm vol}_{S^2}\ ;\\ 
		&e^\phi=2^{5/4}\pi^{5/2} 3^4 \frac{(-\alpha/\ddot \alpha)^{3/4}}{(\dot \alpha^2-2 \alpha \ddot \alpha)^{1/2}}\, .
	\end{align}	
\end{subequations}
Here $\ddot \alpha = \ddot \alpha(\zeta)$ is a piecewise linear function on a closed interval $I$ parameterized by $\zeta$. Flux quantization restricts its derivative to be integer-valued (with the integer related to the Romans mass $F_0$), and its points of discontinuity to be located at integer values of $\zeta$. $\alpha$ is a double integral of $\ddot \alpha$, as the notation implies. The two integration constants can be fixed in various ways. In \cite{cremonesi-t} they were fixed so that $\alpha$ vanishes at the endpoints of $I$. This corresponds to the $S^2$ shrinking smoothly there. Various boundary conditions for $\alpha$ correspond to various physical situations. The cases that occur are:
\begin{itemize}
	\item at an endpoint of $I$ where $\alpha$ and $\ddot \alpha$ have a single zero, and $\dot\alpha^2\neq 2 \alpha \ddot \alpha$, the metric is regular,
	\item at an endpoint of $I$ where $\alpha$ has a single zero, $\ddot \alpha \neq 0$ and $\dot\alpha^2\neq 2 \alpha \ddot \alpha$, there are D6-brane sources,
	\item at an endpoint of $I$ where $\ddot \alpha$ has  a single zero, $\alpha \neq 0$ and $\dot\alpha^2\neq 2 \alpha \ddot \alpha$, there are O6-plane--D6-brane sources,
	\item at an endpoint of $I$ where $\dot\alpha^2- 2 \alpha \ddot \alpha$ and $\ddot \alpha$ have a single zero, there are O8-plane--D8-brane sources. 
\end{itemize}
The first three situations were already identified in \cite{10letter}. The last one we present here and is the AdS$_7$ solution of interest for this paper. It occurs for $b_2=18$ in the formulation of \cite{10letter}. This O8--D8 system is of divergent dilaton type, in the sense discussed after (\ref{eq:c=0}). As noticed there, this is the same that occurs for example in the AdS$_6$ solution of type IIA \cite{brandhuber-oz} supergravity, describing an O8--D8--D4 system. In a sense the solutions of this section are an AdS$_7$ analogue of \cite{brandhuber-oz}. AdS$_7$ solutions with an O8--D8 system and non-divergent dilaton should exist, but are not relevant for the present paper. 

For the solution of interest
\begin{equation}\label{eq:alphaO8}
	\alpha = (3\pi)^3 F_0 (\zeta_0^3-\zeta^3)\, .
\end{equation}
According to the analysis above, at $\zeta=0$ there is an O8--D8 source, while at $\zeta=\zeta_0$ a stack of D6-branes. The presence of the O8-plane means that the solution should be thought of as the ``right half'' of a solution where $\zeta\in [-\zeta_0,\zeta_0]$. The internal manifold for this ``full'' solution would have the topology of a three-sphere $S^3$. However we only consider the right half for which $\zeta\in [0,\zeta_0]$, and correspondingly the internal manifold has the topology of a half-$S^3$.
From (\ref{eq:ads7a}) we obtain 
\begin{subequations}\label{eq:ads7}
	\begin{align}
		&\frac1{\pi \sqrt2} ds^2= 8\sqrt{\frac {\zeta_0^3-\zeta^3}{6\zeta}}ds^2_{{\rm AdS}_7}+ \sqrt{\frac{6\zeta} {\zeta_0^3-\zeta^3}} \left(d\zeta^2 + \frac{(\zeta_0^3-\zeta^3)^2}{3\zeta(4\zeta_0^3-\zeta^3)} ds^2_{S^2}\right)\, ;\\
		&B= \frac{F_2}{F_0}+ \pi \zeta_0{\rm vol}_{S^2}\, ,\qquad F_2 =  \pi F_0 \zeta \frac{2\zeta^3-5\zeta_0^3}{4\zeta_0^3-\zeta^3} {\rm vol}_{S^2}\, ;\\ 
		&e^\phi=\frac{2^{1/2}}{3^{1/4} \pi^{1/2} F_0} \zeta^{-5/4}\frac{(\zeta_0^3-\zeta^3)^{3/4}}{(4 \zeta_0^3-\zeta^3)^{1/2}}\, .
	\end{align}	
\end{subequations}
Applying the map \cite[Eq.~(5.19)]{afpt} (or \cite[Eq.~(5)]{10letter}) to the metric in (\ref{eq:ads7}) we find (\ref{eq:ads5nop}), with $w=\zeta/\zeta_0$ and $z_0 = 3 \pi \zeta_0$. As we anticipated in section \ref{sub:kappa-1}, this provides the link between the AdS$_5$ solutions found in this paper and the AdS$_7$ solutions discussed in this section.

The gauge of $B$ is fixed by demanding regularity at $\zeta=\zeta_0$. Flux quantization (\ref{eq:fluxq}) fixes
\begin{equation}\label{eq:c}
	\frac{\zeta_0}{\ell^2_s} = -\frac{M}{n_0}\,. 
\end{equation}
where 
\begin{equation}
n_0 = 2\pi \ell_s F_0\, , \qquad M = \frac{1}{2\pi \ell_s} \int_{S^2} (F_2 - B F_0)\, .
\end{equation}
$M$ can be interpreted as the number of D6-branes at $\zeta=\zeta_0$. It is also easy to compute the integral of the NS--NS flux $H$: by integration by parts, it reduces to the integral of $B$ over $S^2$ near $\zeta=\zeta_0$ (where it vanishes) and near $\zeta=0$. This gives
\begin{equation}\label{eq:N}
	N= \frac1{(2\pi\ell_s)^2}\int H = \frac{M}{n_0 }\, . 
\end{equation}
This result can be obtained more easily by integrating the Bianchi identity $d F_2 - F_0 H= \delta$, since $F_0$ is constant.

As we remarked earlier, the dilaton diverges near $\zeta=0$. However, this issue is localized in a small region around that locus. Moreover, as we will see, this singularity does not affect the computation of the $a$ anomaly.


\subsection{The field theory interpretation} 
\label{sub:ft}

We will now give the field theory interpretation of the AdS$_7$ solution (\ref{eq:ads7}). 

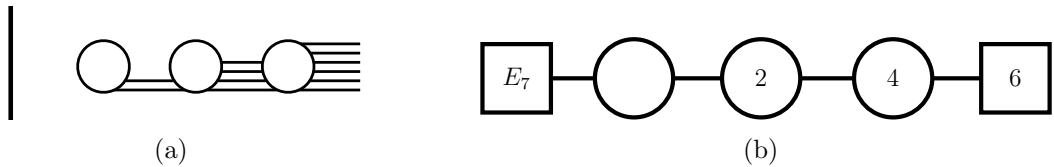
\begin{figure}[ht]
\centering	
	\subfigure[\label{fig:O8-NS5-D6}]{\includegraphics[height=1.5cm]{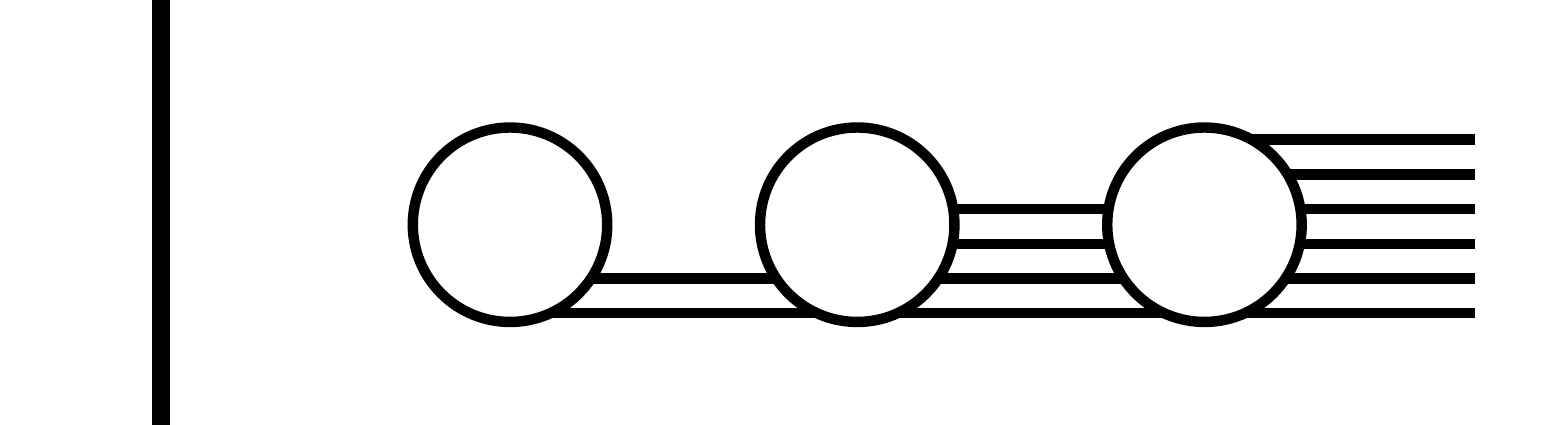}}\hspace{1cm}
	\subfigure[\label{fig:E7-6}]{
	\begin{tikzpicture}[scale=0.65, every node/.style={scale=.8}]

	\draw[fill = white,ultra thick] (-6.7,-.7) rectangle (-5.3,.7);
	\node at (-6,0){$E_7$};

	\draw[ultra thick] (-5.3,0)--(-4.4,0);

	\draw[fill = white,ultra thick] (-3.6,0) circle[radius=8mm];

	\draw[ultra thick] (-2.8,0)--(-1.8,0);

	\draw[fill = white,ultra thick] (-1,0) circle[radius=8mm];
	\node at (-1,0){$2$};

	\draw[ultra thick] (-.2,0)--(.9,0);

	\draw[fill = white,ultra thick] (1.7,0) circle[radius=8mm];
	\node at (1.7,0){$4$};

	\draw[ultra thick] (2.5,0)--(3.5,0);

	\draw[fill = white,ultra thick] (3.5,-.7) rectangle (4.9,.7);
	\node at (4.2,0){$6$};

	\end{tikzpicture}
	}
	\caption{\small In figure \subref{fig:O8-NS5-D6}, the brane diagram whose near-horizon limit produces the solution (\ref{eq:ads7}) for $n_0=2$ and $N=3$ is depicted. The vertical line represents an O8-plane with $n_8=8-n_0$ ($=6$, in this case) D8-brane pairs. The nodes denote NS5-branes, and the horizontal lines D6-branes. In figure \subref{fig:E7-6}, the quiver diagram of the corresponding field theory for this particular case is depicted. The empty node represents the E-string theory, as explained in the main text.}
	\label{fig:ex}
\end{figure}

Following the logic of \cite{gaiotto-t-6d}, we can think of the solution as the result of a near-horizon limit of a brane configuration, which looks like the diagram in figure \ref{fig:O8-NS5-D6}. The vertical line represents an O8-plane with $n_8=8-n_0$ D8-brane pairs, all extended along the directions $(012345789)$. The nodes represent NS5-branes, extended along the $(012345)$ directions. The horizontal lines represent D6-branes extended along the $(0123456)$ directions. Just like in \cite{gaiotto-t-6d}, the idea is that in the near-horizon limit the direction 6 and the radius of the directions $(789)$ mix to produce the radial direction of AdS$_7$ and the coordinate $\zeta$ -- this was recently made more precise in \cite{bobev-dibitetto-gautason-trujien, macpherson-t}. The D6-branes in the brane diagram become the D6-branes at the pole $\zeta=\zeta_0$ in (\ref{eq:ads7}). The $N$ NS5-branes in the brane diagram become the flux integer (\ref{eq:N}).  

Reading off the field theory from the brane diagram is mostly a straightforward application of the standard string theory techniques \cite{brunner-karch, hanany-zaffaroni-6d}. As usual in six dimensions, the field theory one reads off this way is an effective description of the tensor moduli space of a SCFT. The conformal point is really obtained at the origin of this tensor moduli space, which corresponds in the brane diagram to putting all the NS5-branes on top of each other and on top of the O8-plane. 

Having said this, the effective theory is as follows. It consists of a chain of gauge groups ${\rm SU}(i \cdot n_0)$, $i=1,2, \ldots, (N-1)$, coupled to hypermultiplets and tensor multiplets. At the end of the chain there is an ${\rm SU}(N n_0)$ flavor symmetry. (For details, see \cite{brunner-karch, hanany-zaffaroni-6d} or the summary given in \cite[Sec.~2.1]{cremonesi-t}.) The tensor multiplets couple to the gauge fields via a Green--Schwarz--Sagnotti--West mechanism, and via a term of the type $(\phi_i-\phi_{i+1}){\rm Tr}(F_i^2)$, where $\phi_i$ is the real scalar in the $i$-th tensor multiplet and $F_i$ is the $i$-th gauge field strength.

All this is like in the theory corresponding to the NS5--D6 system, shown for example in \cite[Fig.~6]{cremonesi-t}, whose gravity dual is the tear-drop shaped, ``simple massive'' solution of \cite{10letter} and \cite[Sec.~5.2]{afrt}. There is, however, an additional subtlety here, due to the presence of the O8-plane. (This was discussed in section \cite[Sec.~5.1]{hanany-zaffaroni-6d}, although the theories we need are a ``piece'' of the ones in that reference. The theory we are describing appeared recently in \cite{zafrir}.) Via a chain of dualities, the O8--D8 system can be mapped to the $E_8$ wall in M-theory \cite{polchinski-witten}. An unbroken $E_8$ corresponds to a situation where there are 7 D8-brane pairs on the O8-plane; notice that in this situation $n_0 = 2\pi \ell_s F_0$ is equal to 1. When there are fewer D8-brane pairs, $n_8=8-n_0$, the flavor group is broken to $E_{n_8+1}=E_{9-n_0}$. For $n_0=6,5$ this gives $E_7$ and $E_6$ respectively. For $n_8$ smaller still, the sequence was determined in \cite{seiberg-5d} in the context of the five-dimensional CFTs obtained by putting D4-branes next to an O8--D8 system (whose gravity dual \cite{brandhuber-oz} was mentioned earlier):
\begin{equation}\label{eq:En}
\begin{split}
	E_5&= {\rm Spin}(10)\,  ,\qquad E_4 = {\rm SU}(5)\, ,\qquad E_3= {\rm SU}(3)\times {\rm SU}(2)\, ,\\
	E_2&= {\rm SU}(2)\times {\rm U}(1)\, ,\qquad E_1=  {\rm SU}(2)\,.
\end{split}	
\end{equation}

Under the same duality chain, the NS5-branes are mapped to M5-branes. An M5-brane near an $E_8$ M-theory wall is described by a six-dimensional theory with a single tensor multiplet and an $E_8$ flavor symmetry, known as ``E-string theory''. It has a one-dimensional tensor moduli space; conformal symmetry is unbroken at its origin. It also has an F-theory realization, which consists of a single $-1$-curve touching an $E_8$ singularity. This is usually denoted by
\begin{equation}\label{eq:E}
	\lbrack E_8] \,\, {1}.
\end{equation} 
This is actually a rank-1 E-string theory. There is also a rank-$N$ E-string theory, which describes $N$ M5-branes near the $E_8$ wall. This has an $N$-dimensional tensor moduli space. At a generic point, there is an effective description with $N-1$ tensors coupled to (\ref{eq:E}). The F-theory realization consists of a sequence of $-2$-curves ending with (\ref{eq:E}):
\begin{equation}\label{eq:NE}
		\lbrack E_8] \ {1} \ {2} \ldots {2}.
\end{equation}
 
In our setup, the NS5-brane which is nearest to the O8--D8 gives rise to a rank-1 E-string theory (\ref{eq:E}). However, this E-string theory is also coupled to the chain of gauge groups we described earlier, whose first gauge group is ${\rm SU}(n_0)$. The coupling is made by gauging a subgroup ${\rm SU}(n_0)$ of the $E_8$ flavor symmetry of the E-string theory. The leftover flavor symmetry is the commutant of ${\rm SU}(n_0)$ inside $E_8$. 

The algebra of the flavor symmetry can be identified by looking for the maximal subalgebras of $\mathfrak{e}_8$ of the form $\mathfrak{g}\oplus \mathfrak{su}_{n_0}$. One way is to consider semisimple regular subalgebras, which are obtained by deleting a node from the affine $E_8$ Dynkin diagram. The list one obtains this way is 
\begin{equation}\label{eq:regsub}
\begin{split}
	&\mathfrak{e}_7\oplus \mathfrak{su}_2 \,,\qquad \mathfrak{e}_6\oplus \mathfrak{su}_3\, ,\qquad \mathfrak{so}_{10}\oplus \mathfrak{su}_4 \,,\\ 
	&\mathfrak{su}_5\oplus \mathfrak{su}_5 \,,\qquad
	 (\mathfrak{su}_3\oplus\mathfrak{su}_2)\oplus \mathfrak{su}_6 \,,\qquad
	  \mathfrak{su}_2\oplus \mathfrak{su}_8\, .
\end{split}
\end{equation}
We may also consider non-semisimple regular subalgebras, which are obtained by deleting a node from the ordinary (non-affine) $E_8$ Dynkin diagram; the removed dot becomes a $\mathfrak{u}_1$. We get $\mathfrak{su}_2\oplus \mathfrak{u}_1\oplus \mathfrak{su}_7$ this way. We can now obtain the commutant of $\mathfrak{su}_{n_0}$ inside $\mathfrak{e}_8$: for example, the commutant of $\mathfrak{su}_2$ is $\mathfrak{e}_7$, the commutant of $\mathfrak{su}_3$ is $\mathfrak{e}_6$, and so on. This reproduces the full list of ``exceptional'' flavor groups $E_{9-n_0}$ in (\ref{eq:En}), which was originally computed in \cite{seiberg-5d} using heterotic strings. 

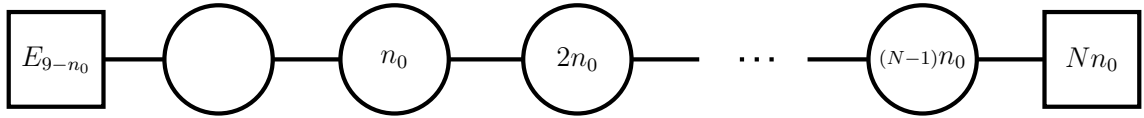
\begin{figure}[ht]
	\centering
		\begin{tikzpicture}[scale=.9, every node/.style={scale=.9}]

		\draw[fill = white,ultra thick] (-6.7,-10.7) rectangle (-5.3,-9.3);
		\node at (-6,-10){$E_{9-n_0}$};

		\draw[ultra thick] (-5.3,-10)--(-4.4,-10);

		\draw[fill = white,ultra thick] (-3.6,-10) circle[radius=8mm];

		\draw[ultra thick] (-2.8,-10)--(-1.8,-10);

		\draw[fill = white,ultra thick] (-1,-10) circle[radius=8mm];
		\node at (-1,-10){$n_0$};

		\draw[ultra thick] (-.2,-10)--(.9,-10);

		\draw[fill = white,ultra thick] (1.7,-10) circle[radius=8mm];
		\node at (1.7,-10){$2n_0$};

		\draw[ultra thick] (2.5,-10)--(3.5,-10);

		\draw[ultra thick, loosely dotted] (4.1,-10)--(4.6,-10);

		\draw[ultra thick] (5.1,-10)--(6.1,-10);

		\draw[fill = white,ultra thick] (6.8,-10) circle[radius=8mm];
		\node at (6.8,-10){$\scriptstyle{(N-1)}$$n_0$};

		\draw[ultra thick] (7.6,-10)--(8.6,-10);

		\draw[fill = white,ultra thick] (8.6,-10.7) rectangle (10.0,-9.3);
		\node at (9.3,-10){$N n_0$};

		\end{tikzpicture}
	\caption{\small The quiver diagram of our six-dimensional theory in the general case.}
	\label{fig:quiver}
\end{figure}

The end result is the theory shown in figure \ref{fig:E7-6} for the brane diagram in figure \ref{fig:O8-NS5-D6}, and in figure \ref{fig:quiver} for the most general case. 

For completeness we also write down the F-theory realization of these theories:
\begin{equation}\label{eq:MEn}
	\lbrack E_{9-n_0}] \
	{1} \hspace{.4cm} 
	\overset{\mathfrak{su}_{n_0}}{2} \hspace{.2cm}
	\overset{\mathfrak{su}_{2n_0}}{2} \hspace{.2cm}
	\ldots
	\overset{\mathfrak{su}_{(N-1)n_0}}{2} \,
	[{\rm SU}(N n_0)].
\end{equation}
The $n_0=1$ case is rather special, because the first gauge group is in fact empty:
\begin{equation}\label{eq:ME8}
	\lbrack E_8] \ 
	{1} \hspace{.35cm} 
	{2} \hspace{.3cm}
	\overset{\mathfrak{su}_2}{2} \hspace{.2cm}
	\overset{\mathfrak{su}_3}{2} \hspace{.2cm}
	\ldots
	\overset{\mathfrak{su}_{(N-1)}}{2} \,
	[{\rm SU}(N)]. 
\end{equation}
These theories are similar to the rank-$N$ E-string theory in (\ref{eq:NE}), except for the presence of the linearly growing gauge groups on the $-2$-curves. They are the analogue of a rank-$N$ E-string theory in the presence of Romans mass $F_0$. For this reason, we will call (\ref{eq:MEn}), (\ref{eq:ME8}) rank-$N$ ``massive E-string theories''.

As a cross-check of our field theory identification, we have computed using the methods in \cite{intriligator-a6, osty-a6}, the leading order behavior of the $a$ anomaly, and compared it against a holographic computation from the gravity dual. For solutions with only D8-branes, which are dual to chains of ${\rm SU}(r)$ gauge groups, this comparison was carried out in \cite{cremonesi-t}, finding perfect agreement for all the infinitely many theories in that class. For the theories discussed in this paper, the computation is a bit different because of the presence of the rank-1 E-string. Recalling the caveat pointed out at the end of section \ref{sub:ads7} about the divergent dilaton, one might be skeptical of a holographic comparison. Indeed for example in \cite{jafferis-pufu}, for the aforementioned AdS$_6$ solution with an O8-plane \cite{brandhuber-oz}, it was found that the on-shell action diverges, hindering comparisons with field theory quantities. In our case, however, the holographic $a$ anomaly is found to be finite, and to reproduce the $a$ anomaly: 
\begin{equation}\label{eq:a6d}
	a\sim \frac{16}7 \frac 9{15} N^3 M^2 
\end{equation} 
at the leading order in the holographic limit; for details, see appendix \ref{app:a}. Notice that (\ref{eq:a6d}) is different, even at leading order, from the theory with linearly growing quivers but without the rank-1 E-string, which was found to have \cite{10letter, cremonesi-t} $a \sim \frac{16}7 \frac 4{15} N^3 M^2$. 



\section*{Acknowledgements}
We would like to thank Ken Intriligator and Vasilis Stylianou for discussions. We thank the organizers of the Simons Center for Geometry and Physics 2015 summer workshop for hospitality at the initial stages of this project. We also thank the organizers of the workshop ``String Theory in London 2016'', where part of this work was completed. The work of I.B.~is supported in part by UC president’s post-doctoral fellowship and in  part  by  DOE  grant  DE-SC0009919. The work of A.P.~and A.T.~is supported by INFN and the European Research Council under the European Union's Seventh Framework Program (FP/2007-2013) -- ERC Grant Agreement n. 307286 (XD-STRING).
A.P.~is also supported by the Knut and Alice Wallenberg Foundation under grant Dnr KAW 2015.0083.
A.T.~was also supported by the MIUR-FIRB grant RBFR10QS5J ``String Theory and Fundamental Interactions'', and by the ERC-STG grant 637844- HBQFTNCER.

\appendix

\section{Comparison with \cite{afpt}} 
\label{app:comparison}

In this appendix we summarize the relation between the presentation of the results of \cite{afpt} in that paper and the one in section \ref{sec:general_system} of the present work. 

The metric of the AdS$_5$ solutions was expressed in \cite[Sec.~3]{afpt} as
\begin{align}
ds^2_{10} &= e^{2W} ds^2_{{\rm AdS}_5} + e^{2\varphi} \left(dx_1^2 + dx_2^2 \right) + \frac{1}{9} b^2 e^{2W} \left(d\psi + \rho\right)^2 + ds^2_{M_2}\,, \\
ds^2_{M_2} &= e^{-4W+ 2 \phi} b^{-2} \left[ \left(b^2 +a_2^2 \right) e^{-4W} dx^2 + \left(b^2 + a_1^2 \right) dy^2 + 2a_1 a_2 e^{-2W} dx dy \right]\,. \label{eq:M2}
\end{align}  
The functions $(W,b,a_1,a_2,\varphi)$\footnote{$W$ is denoted by $A$ in \cite{afpt}.} and the dilaton $\phi$ depend on the coordinates $(x,y, x_1, x_2)$, and obey the relations \cite[Eqs.~(3.6), (3.7)]{afpt}
\begin{equation}\label{eq:a1a2b} 
a_1^2 + a_2^2 + b^2 =1\,, \qquad e^{3W-\phi} a_1 =-2y \,, 
\end{equation}
and \cite[Eqs.~(3.17b), (3.17c), (3.22)]{afpt}
\begin{align}\label{eq:functrel}
3 e^{-3W + \phi} \frac{a_1}{b^2} = \partial_y  \log \left( b e^{\varphi} e^{4W-\phi} \right)\,, \qquad
3 e^{-5W + \phi} \frac{a_2}{b^2} = \partial_x  \log \left( b e^{\varphi} e^{4W-\phi} \right)\,.  
\end{align}  
The one-form $\rho$ is determined by the above functions as \cite[Eq.~(3.17a), (3.23)]{afpt}
\begin{equation}\label{eq:rho}
\rho = -\star_2 d_2 \log \left( b e^{\varphi} e^{4W-\phi} \right)\,,
\end{equation} 
where the Hodge star operator $\star_2$ and the exterior derivative $d_2$ are taken over the $(x_1, x_2)$ plane.  

The transformation relating the coordinates $(s,u)$ of section \ref{sec:general_system} and $(y,x)$ is
\begin{equation}
y^2 = 2s\,, \qquad 3 x = u\,.
\end{equation} 
The functions $(D_s, D_u)$ in that section are defined by the ones appearing in \cite{afpt} as 
\begin{equation}
e^{\widetilde{D}_s} = \frac{12\sqrt{2}}{y^3} b^2 e^{2\varphi} e^{8W-2\phi}\,, \qquad \partial_u D_u = -\frac{1}{3} \left(1+ \frac{a_2^2}{b^2} \right) e^{-4W}\,.
\end{equation}  
Equations \eqref{eq:a1a2b} and \eqref{eq:functrel} are then solved by $(D_s, D_u)$ via the expressions
\begin{align}
b^2   &= -\frac{3}{2s} \frac{1}{\partial_s D_s} \frac{\det(h)}{\det(g)}\,, \qquad 
a_1^2 = \frac{\det(h)}{\det(g)}\,, \qquad
a_2^2 = - \frac{3}{2s} \frac{1}{\partial_s D_s} \frac{(\partial_u D_s)^2}{\det(g)}\, ,
\nonumber \\
e^{4W} &= - \frac{\partial_s D_s}{3 \det(h)}\,, \qquad
e^{2\varphi} = \frac{1}{24} e^{2W} \det(h) e^{D_s}\,, \qquad
e^{2\phi} =  \frac{1}{8s} e^{6W}  \frac{\det(h)}{\det(g)}\,.
\end{align}
See section \ref{sec:general_system} for the definition of $\widetilde{D}_s$, $\det(h)$ and $\det(g)$.
Finally, $\rho =- \frac{1}{2} \star_2 d_2 D_s$, and the metric (\ref{eq:M2}) becomes
\begin{equation}
ds^2_{M_2} = \frac{1}{24 s} \frac{\det(h)}{\det(g)}  e^{2W} \left[-\partial_u D_u\, du^2 - \partial_s \widetilde{D}_s\, ds^2 - 2 \partial_u D_s\, du ds \right]\,.
\end{equation}

\section{D4-branes inside D8-branes}
\label{app:D4inD8}

Solutions describing D$p$-branes inside the wordvolume of D$(p+4)$-branes were constructed in \cite{Youm:1999zs}. Specialising to the case $p=4$, let the D4-branes be extended along the directions $x_0$ to $x_4$ and the D8-branes along all directions except for $x_9$. The spacetime metric of the solution is given by
\begin{equation}
	ds^2 = (H_8 H_4)^{-\frac{1}{2}} (-dx^2_0 + \dots + dx^2_4) + H_4^{\frac{1}{2}} H_8^{-\frac{1}{2}} (dx^2_5 + \dots + dx^2_8) + (H_4 H_8)^{\frac{1}{2}} dx^2_9\,.
\end{equation}
where the functions $H_4$ and $H_8$ satisfy the equations
\begin{equation}\label{D4D8eq}
\partial^2_{x_9} H_4 + H_8 \sum_{i=5}^{8} \partial^2_{x_i} H_4 = 0\,, \qquad
\partial^2_{x_9} H_8 = 0\,.  
\end{equation}
These are solved by
\begin{equation}
	H_4 = 1 + Q_4 \left(r^2+ \frac49 Q_8 |x_9|^3\right)^{-5/3}\,, \qquad H_8 = Q_8 |x_9|\,,
\end{equation}
where $r^2 = \sum_{i=5}^8 (x_i)^2$ is the radial coordinate in the $x_5$ to $x_8$ directions, and $Q_4$, $Q_8$ are constants.

We are interested in a solution where the D4-branes are localized only in the $x_7$, $x_8$ directions and smeared over the $x_5$, $x_6$ ones.
The first of the equations \eqref{D4D8eq} is thus modified as
\begin{equation}
\partial^2_{x_9} H_4 + H_8 (\partial^2_{x_7} + \partial^2_{x_8}) H_4 = 0\,,
\end{equation}
and the $H_4$ that solves it reads
\begin{equation}
	H_4 = 1 + Q_4 \left(r^2+ \frac49 Q_8 |x_9|^3\right)^{-2/3}\,, \qquad 
	r^2 = x^2_7 + x^2_8\,.
\end{equation}

It is convenient to introduce the coordinate $\lambda$ via
\begin{equation}
	|x_9| = \left(\frac49 Q_8 \right)^{-1/3} \lambda^{2/3}\, ,
\end{equation}
and rewrite the spacetime metric as:
\begin{equation}
	ds^2 = \left(\frac{3}{2} Q_8 \lambda\right)^{-1/3} [H_4^{-\frac{1}{2}} (-dx^2_0 + \dots + x^2_4) + H_4^{\frac{1}{2}} (dx^2_5 + \dots + dx^2_8 + d\lambda^2)]\,.
\end{equation}
Near the core of the solution,
\begin{equation}
H_4 \simeq Q_4 \left(r^2+\lambda^2\right)^{-2/3},
\end{equation}
and upon making the coordinate transformation,
\begin{equation}
\lambda = \sigma \cos(\theta), \qquad
r = \sigma \sin(\theta),
\end{equation}
the metric takes the asymptotic form:
\begin{align}
	ds^2 &= \left(\frac{3}{2} Q_8 \sigma \cos(\theta) \right)^{-1/3} [Q_4^{-1/2} \sigma^{2/3} ds^2_{||} + Q_4^{1/2} \sigma^{-2/3} ds^2_{\perp}], \nonumber \\
	ds^2_{||} &= -dx^2_0 + \dots + x^2_4, \qquad 
	ds^2_{\perp} =  dx^2_5 + dx^2_6 + d\sigma^2 + \sigma^2 \left(d\theta^2 + \sin^2(\theta) d\psi^2\right).
\end{align}

\section{The limiting case $z_1=z_0$}
\label{app:z1=z0}

In this appendix, we will analyze the limiting case $z_1 = z_0$ in the space of solutions of section \ref{sub:g0}. For $z_1 \ge z_0$, the solutions contain D6-branes, and an O8-plane with D4- and D8-branes on top. We argued in the main text that they represent gravity duals of a compactification on a two-sphere with punctures of the six-dimensional theories discussed in section \ref{sec:6d}. Here we will see that the limiting case $z_1 = z_0$ is of unclear physical interpretation.

For $z_1=z_0$, $z_0$ becomes a double root of $p$, since $p'(z_1) = 0$. Once this happens, the analysis of section \ref{sub:special_points} is not valid near $z = z_0$ and near $\{z=z_0,\, k=1\}$. Indeed the geometry in these regions departs from the one where $z_0$ is a single root. 

Near $z=z_0$, we introduce the coordinate $\varrho^2 = \frac{8}{3}(1 - z/z_0)$. In the $z \to z_0$ limit the metric becomes
\begin{align}
ds^2_{10} &\sim z_0 \sqrt{\frac{1-k^3}{k}}\left[ds^2_{{\rm AdS}_5} + \frac{kdk^2}{1-k^3} + d\varrho^2 + \varrho^2 \frac{1}{4}\left(\eta_\psi^2 + ds^2_{S^2}\right) \right]\, .
\end{align}
Given that the period of $\psi$ is $2\pi$, $\frac{1}{4}\left(\eta_\psi^2 + ds^2_{S^2}\right)$ represents the metric on a $\mathbb{Z}_2$ orbifold of a three-sphere of unit radius, with the $\mathbb{Z}_2$ acting on the R-symmetry circle. Including the direction parameterized by $\varrho$, we conclude that the internal space contains $\mathbb{R}^4/\mathbb{Z}_2$. 

The locus $\{z=z_0,\, k=1\}$ is of more challenging physical interpretation. In the neighborhood of $\{z=z_0,\, k=1\}$, the local change of coordinates
\begin{equation}
	1 - \frac{z}{z_0} = \frac{3}{8} r^2 \cos^2(\theta)\, \qquad 1-k= \frac{3}{4} r^2 \sin^2(\theta)\,
\end{equation} 
puts the metric in the form 
\begin{align}\label{eq:sq}
ds^2_{10} &\sim \frac{3z_0}{4} r  \sqrt{1+3\sin^2(\theta)} \left(ds^2_{{\rm AdS}_5} + dr^2 + r^2 ds^2_4 \right)\,, \nn \\
ds^2_{4} &= d\theta^2 + \frac{1}{4}\cos^2(\theta)\left(\frac{4\sin^2(\theta)}{1 + 3\sin^2(\theta)} \eta^2_{\psi} + ds^2_{S^2}\right)\,.
\end{align}
This is not reminiscent of a brane singularity familiar to us. Given that this solution arises as a limiting case of a solution with D6-branes present at $\{z=z_0,\, k=1\}$, and the appearance of an $\mathbb{R}^4/\mathbb{Z}_2$ singularity near $z = z_0$, it is tempting to speculate that as $z_r = z_0$ in \eqref{D6limit} becomes a double root, $p'(z_r)$ goes to zero and the base $\Sigma_g = S^2$ shrinks, thus turning the D6-branes into fractional D4-branes, whose near-horizon produces (\ref{eq:sq}).

\section{$a$ anomaly} 
\label{app:a}

In this appendix we will explain how to compute the $a$ anomaly for the massive E-string theories (\ref{eq:MEn}) and (\ref{eq:ME8}), using both field-theoretic and holographic techniques. Part of this computation follows \cite{cremonesi-t}, to which we refer for further details. 

The field theory computation can be performed using the methods of \cite{intriligator-a6, osty-a6}. We will focus directly on the holographic leading order contribution, rather than giving the full detailed computation as in \cite{cremonesi-t}. As in that paper, we use \cite{cordova-dumitrescu-intriligator-a6} $a = \frac{16}7 (\alpha- \beta +\gamma) +\frac67 \delta$, where $\alpha$, $\beta$, $\gamma$ and $\delta$ are the various coefficients in the 't Hooft anomaly polynomial:
\begin{equation}
	I_8 = \frac1{24}(\alpha c_2(R)^2+ \beta c_2(R) p_1 + \gamma p_1(T)^2 + \delta p_2(T))\,,
\end{equation}
where $c_2(R)$ is the second Chern class of the R-symmetry bundle and $p_1(T)$, $p_2(T)$ are the first and second Pontryagin classes of the tangent bundle respectively. The holographic limit consists in taking the number $N$ of gauge groups to be large. The limit can be taken in such a way as to leave the supergravity solution essentially unchanged; see \cite[Sec.~2.2.4]{cremonesi-t}.
The leading order contribution comes from $\alpha$, the coefficient of $c_2(R)^2$. This in turn comes from a Green--Schwarz--Sagnotti--West mechanism. Let us first explain how this worked in \cite{cremonesi-t}, and then how it is modified for the case of this paper. The relevant terms of the anomaly polynomial read
\begin{equation}\label{eq:I}
	I= - \frac{1}{8} \sum_{i,j}C_{ij}{\rm tr}F_i^2 {\rm tr}F_j^2 - \frac{1}{2} r_i c_2(R) {\rm tr}F_i^2 + \ldots \,,
\end{equation}
where $C_{ij}$ is the Cartan matrix of $A_{N-1}$. Gauge anomalies should be canceled: this leads us to postulate the presence of a Green--Schwarz--Sagnotti--West \cite{green-schwarz-west,sagnotti} mechanism which gives a further contribution $\frac{1}{8}\sum_{i,j}C_{ij}I_i I_j$, $I_i = {\rm tr}F_i^2+2 C^{-1}_{ij}r_i c_2(R) $. This cancels the two terms appearing in (\ref{eq:I}), but introduces the term $\frac{1}{2} \sum_{i,j}C^{-1}_{ij}r_i r_j c_2(R)^2$. 
This is the leading contribution to $\alpha$ and hence to $a$:
\begin{equation}
	a\sim \frac{192}7 \sum_{i,j}C^{-1}_{ij}r_i r_j  \,.
\end{equation}
If we consider the theory with linearly growing ranks, $r_i=i \cdot n_0$ \cite[Fig.~6]{cremonesi-t}, we can compute $\sum_{i,j}C^{-1}_{ij}r_i r_j = \frac1{180}N(1-N^2)(1-4N^2)\sim \frac1{45}N^5$, which gives
\begin{equation}\label{eq:alin}
	a \sim \frac{16}7 \frac4{15}N^5n_0^2 = \frac{16}7 \frac4{15}N^3 M^2 \, ,
\end{equation} 
where equation (\ref{eq:N}) is used.

In our case, the computation is modified by the presence of an E-string. The relevant part of its anomaly polynomial is 
\begin{equation}\label{eq:IE}
	I_{\text{E-string}} = -\frac{1}{4}{\rm Tr}F^2 c_2(R) + \frac{1}{32} ({\rm Tr}F^2)^2 \,.  
\end{equation}
Here $F$ is the $E_8$ field strength, and ${\rm Tr}\equiv \frac1{30}{\rm tr}_{\rm fund}$. There is also a ${\rm Tr}F^2 p_1(T)$ term which would contribute to $\beta$ and $\gamma$, but their contributions to $a$ are subleading. When an ${\rm SU}(n_0)$ subgroup of $E_8$ is gauged, (\ref{eq:IE}) modifies (\ref{eq:I}) in two ways: 
\begin{equation}
	C_{ij}\to \tilde C_{ij}= C_{ij}- \delta_{i1} \delta_{j1} \, ,\qquad r_i \to r_i - \delta_{i1}\, .
\end{equation}
Of these two, only $C\to \tilde C$ affects the leading order. Hence, we end up with 
\begin{equation}
	a\sim \frac{192}7 \sum_{i,j}\tilde C^{-1}_{ij}r_i r_j \,.
\end{equation}
In our case, again, $r_i = i \cdot n_0$. We evaluate $\sum_{i,j}\tilde C^{-1}_{ij}r_i r_j= \frac1{60}N(N+1)(N+2)(3N^2 + 6 N +1)\sim \frac1{20}N^5$, which finally leads to 
\begin{equation}\label{eq:aus}
	a \sim  \frac{16}7 \frac 9{15}N^3 M^2 \,.
\end{equation}

We will now compare the above result with a holographic computation. As argued in \cite[Sec.~4]{cremonesi-t}, the latter reduces to 
\begin{equation}\label{eq:ahol}
	a_{\rm hol} = \frac{192}{7 \cdot 3^8 \pi^4\ell^8_s}\int \alpha \ddot \alpha \, d\zeta\, .
\end{equation}
where $\alpha = \alpha(\zeta)$ is the function that characterizes the supergravity solution in (\ref{eq:ads7a}).  As explained in \cite{cremonesi-t}, this is a continuum version of $\sum_{i,j}r_i C^{-1}_{ij}r_j$, keeping in mind the fact that $C_{ij}$ is a ``discrete double derivative''. For example, the gravity solution corresponding to the chain of gauge groups SU$(i \cdot n_0)$, $i=1,\ldots, (N-1)$ is the so-called ``simple massive'' solution. In the present language, it is given by (\ref{eq:ads7a}) with 
\begin{equation}
	\alpha =  (3\pi)^3 F_0 \zeta (\zeta_0^2-\zeta^2)\, ,
\end{equation} 
Performing the integral (\ref{eq:ahol}) one reproduces (\ref{eq:alin}). 

On the other hand, for the rank-$N$ massive E-string theories, we should use $\alpha$ as in (\ref{eq:alphaO8}). This reproduces (\ref{eq:aus}). 

Notice that the two solutions we just considered have the same $\ddot \alpha$, and only differ by its double indefinite integral $\alpha$. In \cite{cremonesi-t}, this integral was fixed by demanding it to vanish at the extrema. This, on the field theory side, corresponds in a way to the choice of $C_{ij}$ rather than $\tilde C_{ij}$ as a discrete version of the double derivative. 

We expect that the match obtained in this section works for more general quivers with an E-string at their end. However, we will not demonstrate this here, since such theories are for the time being not relevant to generating AdS$_5$ solutions similar to the ones studied in this paper.

\bibliography{at}
\bibliographystyle{utphys}

\end{document}